\documentclass[aps,prb,twocolumn,citeautoscript,superscriptaddress,nofootinbib,eqsecnum,10pt]{revtex4-2}   

\pdfoutput=1

\usepackage{amsmath,amssymb,bm} 
\usepackage{graphicx}
\usepackage[tight]{subfigure}    
\usepackage{color} 
\usepackage[papersize={8.5in,11in}]{geometry}
\usepackage[colorlinks=true]{hyperref}  
\hypersetup{
	bookmarks=true,         
	unicode=false,          
	pdftoolbar=true,        
	pdfmenubar=true,        
	pdffitwindow=false,     
	pdfstartview={FitH},    
	pdfkeywords={keyword1} {key2} {key3}, 
	pdfnewwindow=true,      
	colorlinks=true,       
	linkcolor=magenta, 
	citecolor=blue,        
	filecolor=magenta,      
	urlcolor=blue           
} 

\usepackage{amsfonts}
\usepackage{slashed}
\usepackage{latexsym}

\usepackage{scrlayer-scrpage}			
\usepackage{graphicx}			

\usepackage{amsfonts}
\usepackage{amsthm}

\usepackage{xcolor}
\usepackage{hyperref}

\begin{document}

\title{Symmetry-breaking signatures of multiple Majorana zero modes in one-dimensional spin-triplet superconductors}
\author{Arnab Barman Ray}
\affiliation{The Institute of Optics, University of Rochester, Rochester, 14623, USA}
	
\author{Jay D. Sau}
\affiliation{Department of Physics, Condensed Matter Theory Center and Joint Quantum Institute, University of Maryland, College Park, MD 2074}
	
\author{Ipsita Mandal}
\affiliation{Laboratory of Atomic And Solid State Physics, Cornell University, Ithaca, NY 14853, USA}

\begin{abstract}
We study the effects of various symmetry-breaking perturbations on the experimentally measurable signatures (such as conductance and Josephson response) of quasi-one-dimensional (quasi-1D) spin-triplet superconductors. In the first part of the paper, we numerically compute the zero and nonzero temperature conductances of the quasi-1D nanowires that host multiple Majorana zero modes. Following the discussion of the case of s-wave Rashba nanowires, we shift to the main focus, i.e., multichannel spin-triplet superconductors. Applying gate voltages (which changes the symmetry of the spin-orbit coupling) as well as magnetic fields to the nanowire, tunes the system between different symmetry classes by splitting the multiple Majorana zero modes. We study how the conductance tracks the topological invariants and the spectra in all these cases. In the second part of the paper, we study the effects of the symmetry-induced spectrum-breaking on the Andreev spectra of Josephson junctions. Similar to the case of the conductance studies, we find that the spectrum shows multiple zero-energy Andreev bound states in the highly symmetric case with mirror and chiral symmetries.
\end{abstract}

\maketitle
	
\tableofcontents

	
\section{Introduction }
\label{intro}
		
Majorana zero modes (MZMs) in topological superconductors~\cite{m1,m2,m4,m5,m6} have generated significant interest because of their potential 
utility in topological quantum computation~\cite{nayakrmp}. For this 
purpose, proximity-induced topological superconductors based on 
semiconductors~\cite{fu-kane,*sau,*lutchyn,*oreg} have 
been considered particularly convenient, because of the large tunability 
resulting from the conventional nature of the constituents. 
However, quasi-one-dimensional (quasi-1D) topological superconductors with the 
potential for harboring multiple MZMs, while not ideal for quantum 
computation applications, are interesting systems in their own right. 
According to the classification table for topological systems~\cite{kitaev-classi,*Schnyder}, one-dimensional (1D) superconductors 
can support Kramers pairs of Majoranas or multiple Majoranas, 
where the systems are time-reversal symmetric (class DIII) \cite{kane-mele}
or chiral symmetric (class BDI) \cite{fidkowski,ipsita-sudip,Mandal2015,*Mandal2016a,*Mandal2016b}, respectively. 
While proximity effect in wide semiconductor nanowires can lead to 
multiple Majoranas in class BDI for the appropriate spin-orbit coupling~\cite{Tewari_2011}, class DIII Majorana Kramers pairs are found to require 
interactions to generate from spin-singlet proximity effect~\cite{jelena,*ips-kramers}.   
Multiple MZMs have turned out to be particularly interesting because of 
novel phenomena that can result from their interplay with interactions. 
The most direct addition of interactions in this case was shown to modify 
the $Z$ invariant to $Z_8$~\cite{fidkowski}. Recently, more interesting physics has been shown to arise from the interplay of such multiple MZMs with random interactions in the form of the Sachdev-Ye-Kitaev (SYK) model~\cite{sy,*kitaev,*marcel}. From a more pedestrian standpoint, details of experimental manifestations, such as quantization 
of conductance or degeneracy of Josephson spectra, are expected to be 
more intricate for systems with multiple MZMs as compared to the ones 
with single MZMs (that have dominated experimental attention so far).
Specifically, it has been shown~\cite{beenakker} that the conductance into a 
wire in the BDI symmetry class takes values that are integer multiples of the 
quantum of conductance. Furthermore, perturbations that reduce the 
symmetry to class D, also reduce the conductance to the characteristic single 
quantum of conductance or vanishing conductance, associated with the 
$Z_2$ topological invariant. 

\begin{figure}
\includegraphics[scale=0.1]{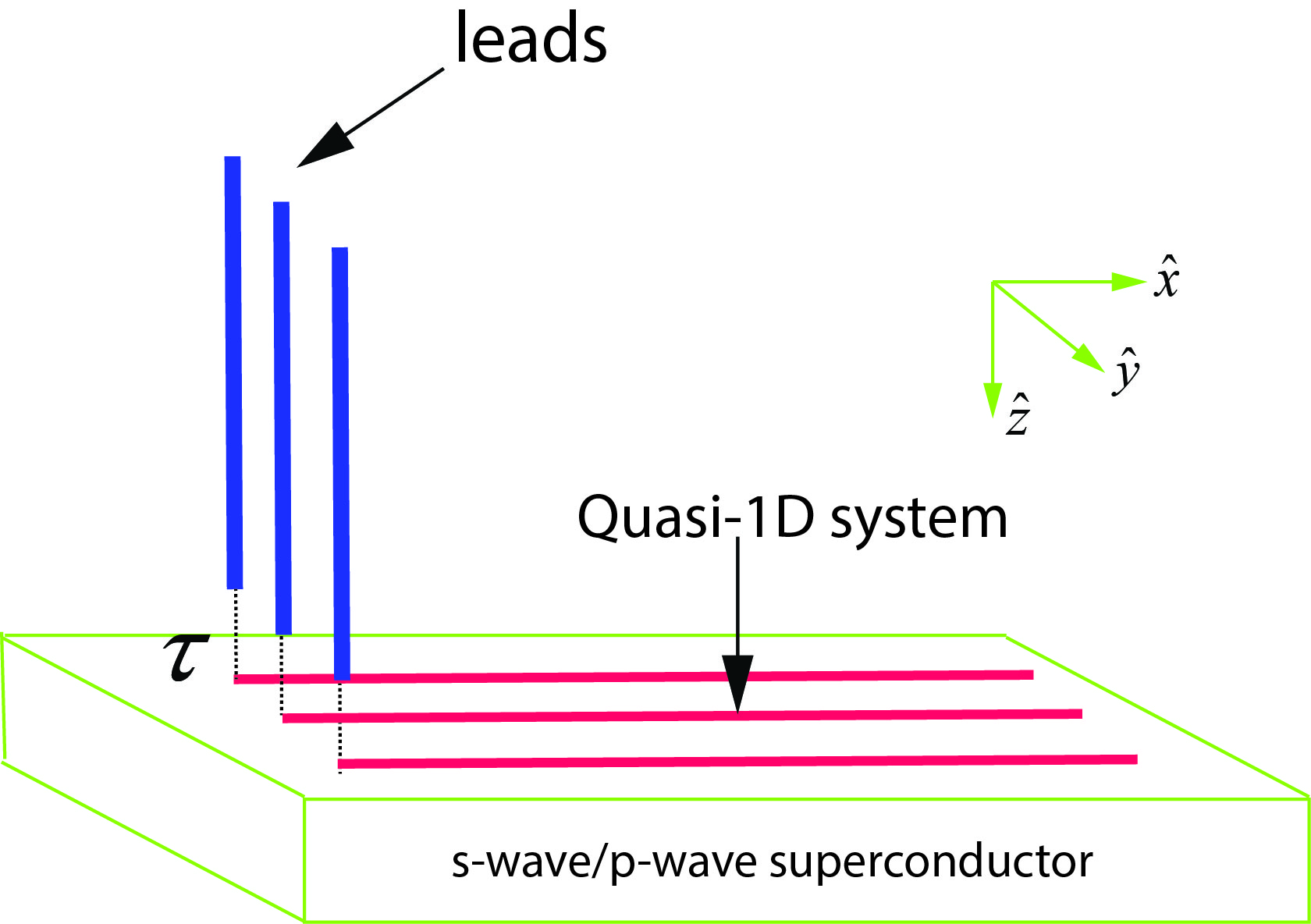}
\caption{\label{gs}
Schematic representation of the generic system treated in Sec.~\ref{tomo}. The strength of the tunnelling barrier is fixed at $\tau=5\,t_x$, where $t_x$ is the hopping along the chain direction. The superconducting system, as well as the leads, are quasi-1D in nature ($t_y << t_x$).}
\end{figure}	
	
Quasi-1D superconductors, that may be viewed  as 
weakly coupled 1D chains \cite{Kaladzhyan_2017} (as shown in Fig.~\ref{gs}), have been suggested in several potential spin-triplet superconductors, such as lithium 
purple bronze~\cite{limo1,Lebed_13}, Bechgaard salts~\cite{Lebed_00,triplet2,Shinagawa_2007}, and even possibly SrRuO$_4$
~\cite{raghu}. Given the evidence for spin-triplet pairing in these systems, 
in the form of high upper critical fields, these materials have been conjectured to 
host MZMs at the ends~\cite{Yakovenko}. Their quasi-1D 
structure, composed of many chains coupled by weak transverse 
hopping, suggests the possibility of one MZM from each of the chains. 
Recent work shows~\cite{Dumitrescu13} that in the cases of time-reversal 
(TR) invariant superconductivity in the form of equal spin pairing (ESP), 
these MZMs would not split, leading to the possibility of multiple MZMs 
at the ends of these materials. Furthermore, spin-triplet superconductors can support
low-energy end-modes even in the absence of external perturbations such as 
magnetic fields. This is different from topological superconductivity in 
semiconductor systems~\cite{*sau,*lutchyn,*oreg}, which despite being 
topologically equivalent to $p$-wave superconductors in certain limits, 
cannot realize time-reversal invariant phases without magnetic fields.
 The flexibility of p-wave systems to realize highly symmetric MZMs, allows for topological superconductivity with a high degree
of symmetry. As shown earlier~\cite{Dumitrescu13}, the stability of the multiple MZMs depends on the variety of symmetries of the systems, and therefore, in principle, can be split by a variety of perturbations.

In this paper, we study the effect of various symmetry-breaking 
perturbations on experimentally measurable signatures of the quasi-1D 
spin-triplet superconductors, such as transport and Josephson response~\cite{yukio1,yukio2,jorge1,jorge2}. 
The fact that pristine quasi-1D spin-triplet superconductors can host MZMs with both time-reversal (class DIII) as well as chiral symmetry (class BDI) provides a rich play-ground for interaction of multiple MZMs with different symmetries. External perturbations such as electric and magnetic fields can couple the MZMs in various ways. Our study of the effect of these perturbations on transport will lead to predictions that experimentalists can use to establish these systems as a platform for controlled multi-MZM systems.
In the first part of the paper (Sec.~\ref{tomo}), we numerically compute the zero and non-zero temperature conductances of quasi-1D nanowires, that host multiple MZMs 
in the configuration depicted in Fig.~\ref{gs}. For the purposes of reference, we start by reviewing the results~\cite{beenakker} on the conductance of the quasi-1D s-wave Rashba nanowire, with parameters chosen such that the system is in the BDI symmetry class. In this case, we study how the conductance into the wire, as a function of density, tracks the bandstructure and the topological invariant -- it is shown to decouple into single nanowires with modified chemical potentials that belong to the BDI class. Following 
this (Sec.~\ref{pwave}), we shift to the main focus, i.e., spin-triplet superconductors. 
We extend the class of perturbations previously considered~\cite{Dumitrescu13}, 
and start with a model with mirror, chiral, and TR symmetries. We systematically break these symmetries by changing various spin-orbit coupling 
terms that maybe controlled by gate voltages and magnetic fields. We study how 
the conductance tracks the topological invariant and spectrum in all these cases.
In the second part of the paper (Sec.~\ref{cavity}), we study the effect of the symmetry-induced spectrum-breaking on the 
Andreev spectra of Josephson junctions (JJs)~\cite{Dumitrescu13}. Recent 
measurements have demonstrated the ability to measure aspects of the Andreev 
state spectrum in a JJ, by two-tone spectroscopy~\cite{urbina,Devoret}. Similar to the case of conductance, we show that the spectrum shows 
multiple zero-energy Andreev states in the highly symmetric case with mirror and 
chiral symmetries.

\section{Differential conductance with normal leads}
\label{tomo}

In this section, we analyze the behaviour of the differential conductance, that can be detected using normal leads connected to the first lattice sites of the system, as shown in Fig.~\ref{gs}. Let $t_{x}$ be the hopping strength between the neighboring sites in the same chain, and $t_{y}$ be the hopping strength between the neighboring sites in the neighboring chains. We consider the limit $t_y\ll t _x$ in order to model a quasi-1D chain.
The leads are modeled as having only hopping (of strength $t_x$) and chemical potential ($\mu_j$) terms corresponding to the single chains, and they are represented by the Hamiltonian
	\begin{align}
H_{leads} & =
-t_{x}  \sum_{i=1}^{N_\ell-1} \sum_{j=1}^{N_y}
	 {\Phi}^{\dagger}_{i+1, j} \,\tau_{z}\, {\Phi}_{i,j} 
\nonumber \\ &
\hspace{1 cm}	 
-  \left(  \tau \,t_{x} \sum_{j=1}^{N_y}
	 {\Phi}^{\dagger}_{1, j} \,\tau_{z}\, {\psi}_{1,j} +  {\rm h.c.}  \right) .
\label{eqlead}	 
	 \end{align}
Here, $\tau $ is the strength of the tunnelling barrier in units of $t_x$, and is set at $5$ for the systems we analyze. Furthermore, ${\Phi}^{\dagger}_{i,j}= (d^{\dagger}_{i,j,\uparrow},d^{\dagger}_{i,j,\downarrow},d_{i,j,\downarrow},- d_{i,j,\uparrow})$ and
${\psi}^{\dagger}_{\ell,j}= (c^{\dagger}_{\ell,j,\uparrow},c^{\dagger}_{\ell,j,\downarrow},c_{\ell,j,\downarrow},- c_{\ell,j,\uparrow})$ are the spinors belonging to the lead and chain sites, respectively.
The site-indices $(\ell,j)$ label the fermions in the $(x,y)$ strip, such that $\ell \in [1,N_x]$ and $j \in [1,N_y]$.
Lastly, $ {\sigma}_\alpha $ and $ {\tau}_\alpha $  ($ \alpha \in \lbrace x,y,z \rbrace $) are the Pauli matrices which act on the spin and particle-hole spaces, respectively . In our numerics, we have taken the number of sites in each chain (lead)
to be  $N_x \, (N_\ell)  =  100$, while the number of chains is set to $N_y =3$, corresponding to a three-channel lead. Energy eigenvalues, voltages, and all the parameters in the Hamiltonian are expressed in units of $t_x$. The conductance is calculated in units of $\frac{e^2}{h}$.
	
 The zero temperature ($T=0$) conductance $G_0(V)$ has been computed using the KWANT package \cite{Groth_2014}, that uses the scattering matrix formalism. These results are extended to  non-zero temperature ($T>0$) conductance using the convolution:
	\begin{equation} 
	G_T (V) = \int dE\, \frac{df_{\epsilon}(T)}{d\epsilon} \Big \vert_{(E-V)} G_0(E)\,,
\label{GT}
	\end{equation} 
	where $f_{\epsilon} (T)=\frac{1}{e^{\epsilon/kT}+1}$ is the Fermi function at temperature $T$ and energy $\epsilon$, whose derivative with respect to energy is carried out at the value $(E-V)$, energy being in units of electron volts.

\subsection{Rashba nanowire}
\label{swave}	

\begin{figure*}[]
{\includegraphics[width = 0.5 \textwidth]{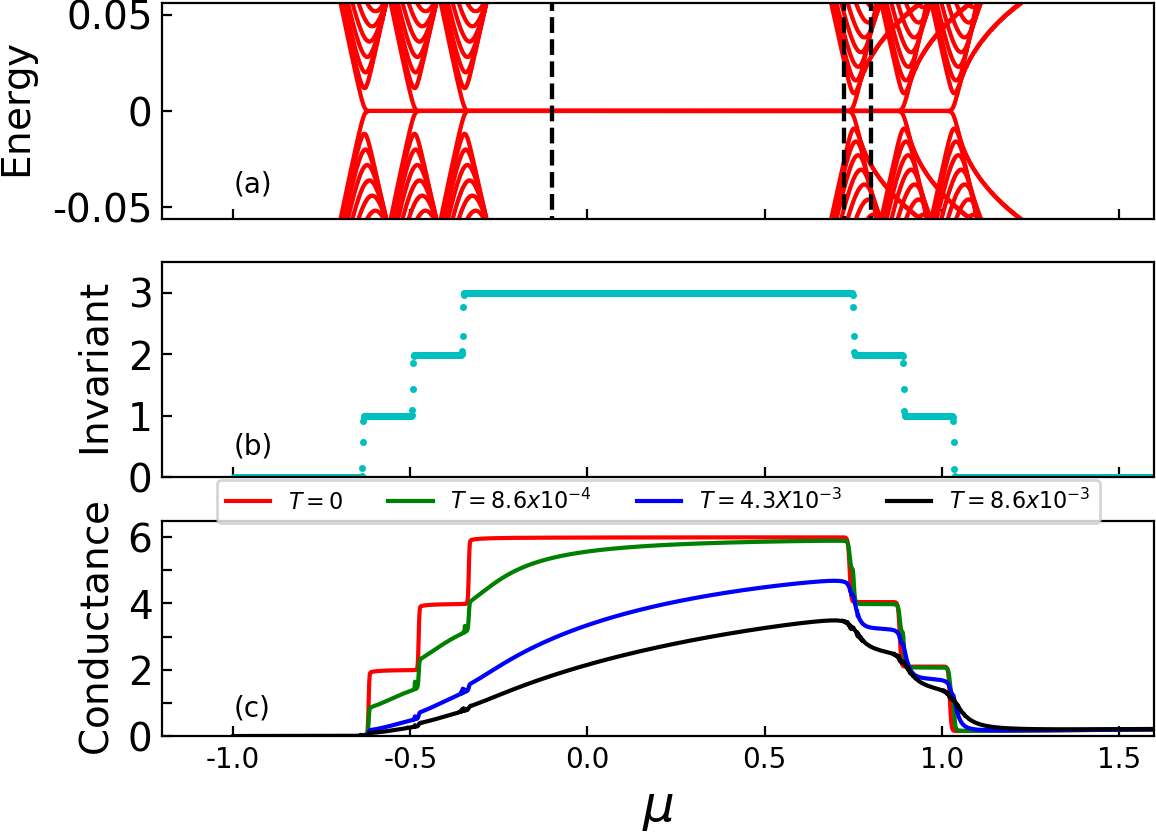}} \hspace{1.5 cm}
{\includegraphics[width = 0.35\textwidth,height = 0.35\textwidth]{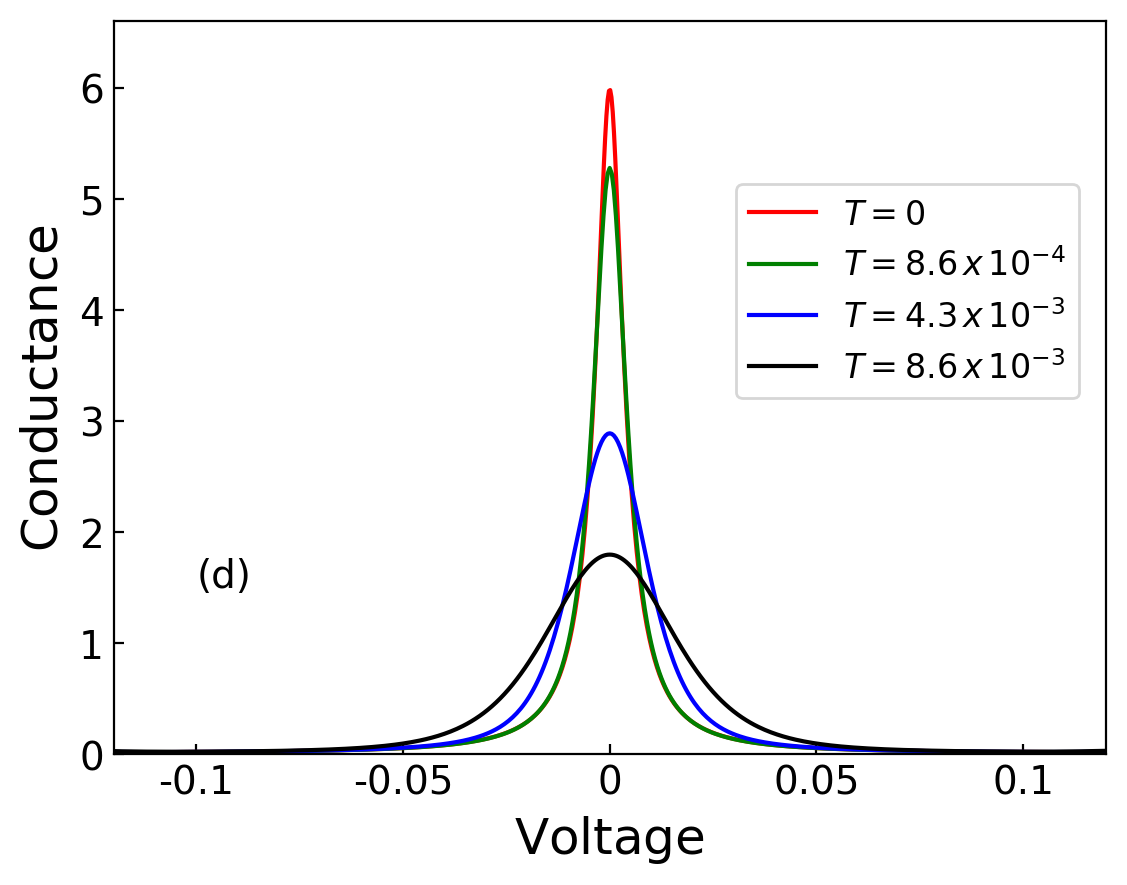}}\\
\hspace{ 1 cm}
{\includegraphics[width = 0.35\textwidth,height = 0.35\textwidth]{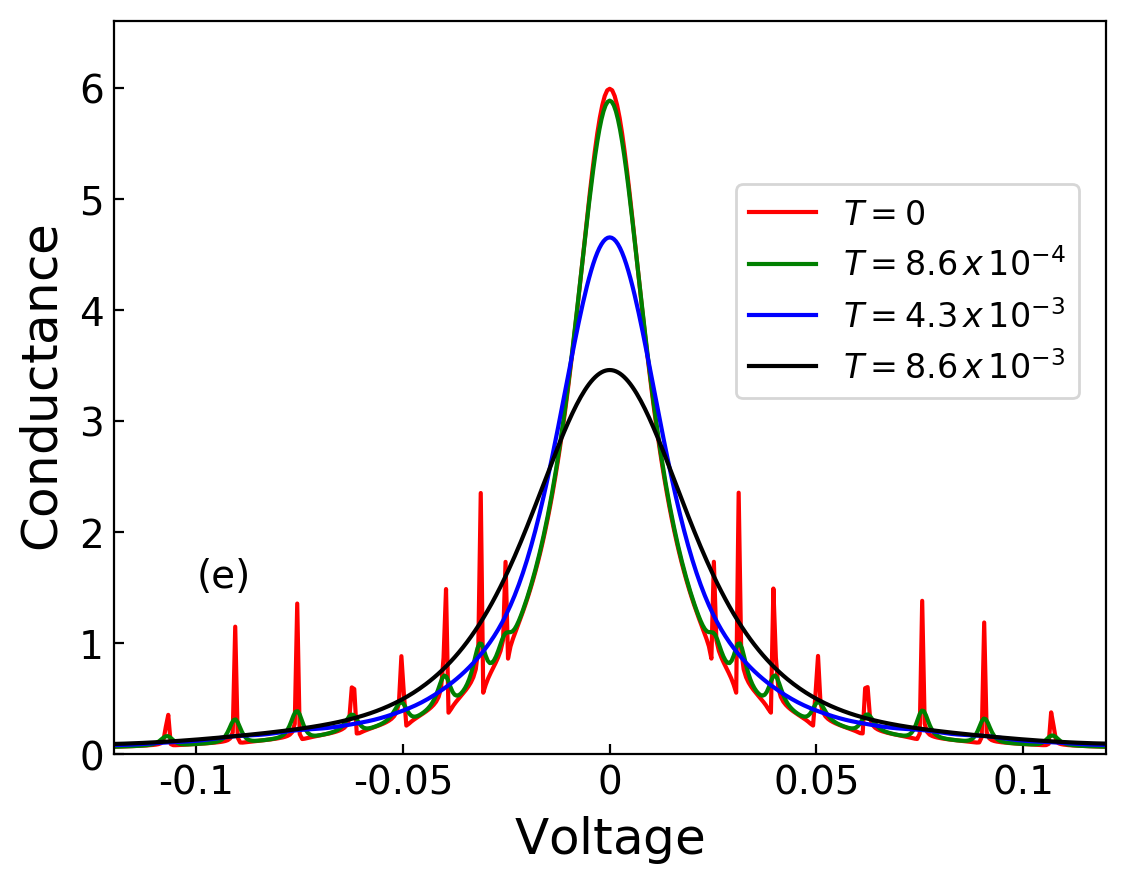}}
\hspace{ 3 cm}
{\includegraphics[width = 0.35\textwidth,height = 0.35\textwidth]{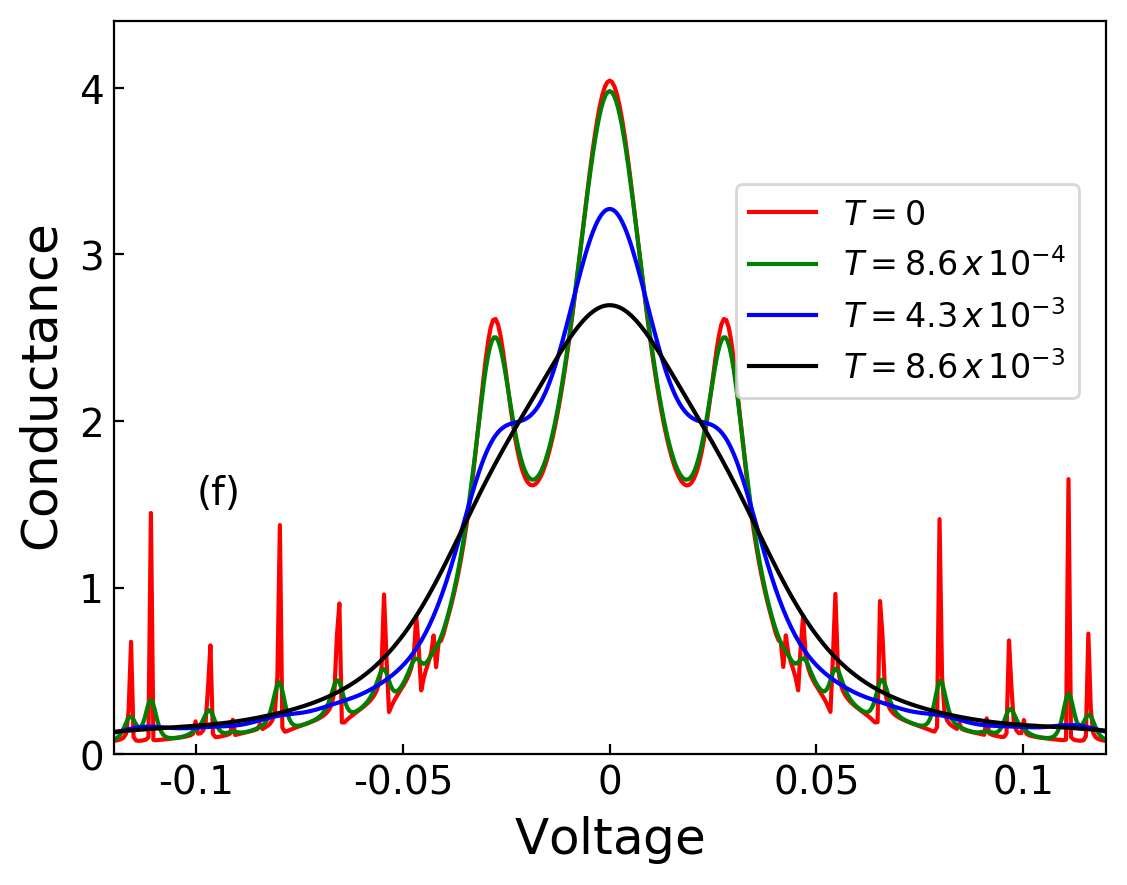}}
\caption{\label{1dnano}
We shows the results for the quasi-1D Rashba nanowire described by Eq.~\ref{nanoham}, with $\Delta_s=0.4,\, V_x=0.8,\,\alpha=0.25, \,,t_y=0.1 $. The first panel (including (a), (b), and (c)) shows the correspondence between the spectrum, the invariant, and the zero-bias differential conductance, all plotted as functions of the chemical potential $\mu$. The remaining three panels (i.e., (d), (e), and (f)) show the differential conductance as a function of voltage, for $\mu=-0.1,\,0.725,$ and $0.8$ respectively, at different temperature values (as shown in the plot-legends).
The black dashed vertical lines in (a) correspond to these three values of $\mu$.
All the temperature values are in units of $t_x$.}
\end{figure*}

We consider a multichannel 1D Rashba nanowire aligned along the $x$-axis, brought in contact with an s-wave superconductor, in the presence of an external magnetic field of strength $V_x$ applied in the $x$-direction. This can be modeled as an array of 1D chains coupled by a weak hopping amplitude, with the Hamiltonian written below:
\begin{align}
\label{nanoham}
H = & \sum_{\ell=1}^{N_x}\sum_{j=1}^{N_y} {\psi}^{\dagger}_{\ell,j}
\left [
\left \lbrace -\mu+2 \left (t_x+t_y \right ) \right \rbrace \tau_{z}
+\Delta_s \tau_{x}+V_{x} \sigma_{x} \right ] {\psi}_{\ell,j}
\nonumber \\	
&	- \sum_{\ell=1}^{N_x-1} \sum_{j=1}^{N_y}
	\left \{ {\psi}^{\dagger}_{\ell+1, j}
\left (t_{x}+ \mathrm{i}\,\alpha\, \sigma_{y}\right )\tau_{z}\, {\psi}_{\ell,j}+{\rm h.c.}  \right\}
\nonumber \\	&
- \sum_{\ell=1}^{N_x} \sum_{j=1}^{N_y-1}
	\left ( {\psi}^{\dagger}_{\ell, j+1} \,t_{y}\,\tau_{z}\, {\psi}_{\ell,j} 
+{\rm h.c.} \right) .
\end{align}
Here, $\mu$ is the chemical potential, $\Delta _s$ is the magnitude of the s-wave superconducting gap, and $\alpha $ is the spin-orbit coupling. 
The system has a chiral symmetry operator $\mathcal{S} = \sigma_y\, \tau_y$, which can used to off-block diagonalize the Hamiltonian~\cite{tewsau} to the form $\begin{pmatrix} 0 & A(k) \\ A^{T}(-k) & 0\end{pmatrix}$. The topological invariant is calculated as~\cite{tewsau}:
\begin{equation} \label{e1} 
\mathcal{Z} = \frac{1}{2 \,\pi \, \mathrm{i}}\,\int_{k=-\pi}^{k = \pi}d\theta(k) \,,\quad 
\theta(k)= \frac{\det \left (A(k) \right)}{|\det \left (A(k) \right )|}\,.
\end{equation}
More details of this calculation are presented in the Appendix. 

The topological behaviour of the quasi-1D Rashba wire described by 
Eq.~\ref{nanoham} can be understood from the evolution of the spectrum of 
the nanowire with a Zeeman field, as shown in Fig.~\ref{1dnano}(a). 
The gap-closures seen in this spectrum indicate the presence of a sequence of six  
topological phase transitions. In addition, for chemical potential in the range $0.75\lesssim\mu\lesssim1.5$, one can see Andreev bound states \cite{DEGENNES,PhysRevB.86.100503,PhysRevB.86.180503,PhysRevB.97.165302,PhysRevB.96.075161,PhysRevB.98.155314,Vuik_2019} as states that have ``peeled off'' below the continuum of states. These states are similar to those obtained in single channel nanowires in the absence of any lead quantum dot, and have an energy that approaches zero at the phase transitions~\cite{PhysRevB.86.100503,PhysRevB.86.180503,PhysRevB.97.165302,PhysRevB.96.075161,PhysRevB.98.155314,Vuik_2019}.

The topological phase transitions seen in Fig.~\ref{1dnano}(a) are not accompanied by a change in symmetry. Rather, they correspond to changes in the topological invariant that is calculated using Eq.~\ref{e1}, and plotted in Fig.~\ref{1dnano}(b). We see that the topological invariant changes between consecutive integer values. As the chemical potential crosses each phase transition, we observe the corresponding gap closures in (a), as expected for topological phase transitions. The range of integer values (i.e., from zero to three) for the topological invariant in this system can be understood from the fact that the normal state (i.e., $\Delta_s=0$) of the isolated system can be decomposed into a sequence of three sub-bands with different wave-function profiles in the $y$ direction. Adding superconductivity does not couple these bands, and hence the Hamiltonian in Eq.~\ref{nanoham} describes a stack of decoupled topological nanowires, with their normal state bands shifted relative to each other. The applied Zeeman field splits the spin components of each of these sub-bands in a way such that they are topological in a range of chemical potentials, where one of these spin-split bands is occupied. Because the Zeeman splitting of the electrons for $V_x=2$
is larger than this separation between the various sub-bands, 
changing chemical potential can sequentially drive all of them into a topological phase,
before the lowest sub-band gets both its spins occupied. This leads to a situation where 
the number of topological bands can increase to three before decreasing back to zero
\cite{PhysRevB.84.144522}. While the results for the topological invariant in this sub-section can be understood from the sub-band decomposition of the Hamiltonian in Eq.~\ref{nanoham}, in later sub-sections we will find that the presence of chiral symmetry protects the topological invariant from small perturbations that couple the sub-bands. 

While the topological invariant, plotted in Fig.~\ref{1dnano}(b), is not directly measurable in experiments, one can see the evidence for this invariant in the zero-bias conductance, plotted in Fig.~\ref{1dnano}(c).
We observe that the zero temperature zero-bias conductance tracks the integer topological invariant in (b) quite closely. The quantization of the conductance here represents the topological invariant, despite the fact that the tunneling Hamiltonian from the lead (Eq.~\ref{eqlead}) couples the different channels, which we used in the previous paragraph to understand the integer values. Thus, measuring the zero-bias conductance at low-enough temperatures can provide insights into the topological phase diagram of such a wire.
Furthermore, the Andreev bound states (ABSs) do not appear to affect the value of the zero-bias conductance peak (ZBCP) for the parameters of our calculation. Fig.~\ref{1dnano}(c) tracks the change of the conductance with a rise in temperatures. We find that the conductance values are lowered as the temperature is increased. However, even at $T =4.3 \times 10^{-3}\,t_x$, the phase transition points can be identified from the conductance plots, though the conductance is significantly reduced from the correct quantized value. 

The thermal suppression of the zero-bias conductance, as seen in Fig.~\ref{1dnano}(c), can be understood by considering the conductance as a function of voltage for different temperatures. One sees from Fig.~\ref{1dnano}(d) (plotted for $\mu=0.1$) that the temperature suppression arises from a broadening of the zero-bias peak with temperature. The conductance as a function of voltage at two other values of $\mu$ (viz. $\mu=0.725$ and $0.8$), plotted in Figs.~\ref{1dnano}(e) and \ref{1dnano}(f), shows a similar thermal suppression.
Interestingly, this thermal-suppression-effect appears to be weaker in the case of the smaller conductance peaks, which are associated with fewer MZMs. This is accompanied by narrower zero-bias peak widths for the case of larger number of MZMs. These observations suggest that the extra MZMs, that occur in the case of larger number of modes, are coupled to the lead with a weaker tunneling amplitude.  
This behaviour is found in all the cases examined in this paper. The $T=0$ plot in Fig.~\ref{1dnano}(e) shows sharp peaks away from the zero voltage, which are associated with the bulk states that are quantized by finite size effects. The width of these resonances are suppressed because of the weak tunneling 
of these states across the tunnel barrier. These narrow peaks associated with the
sub-gap states are washed away at higher temperatures.
Furthermore, Fig.~\ref{1dnano}(f) shows additional broad peaks, away from zero-energy but below the superconducting gap. These peaks are associated with finite energy ABSs, the evidence of which is seen in the spectrum in Fig.~\ref{1dnano}(a).
The ABSs that result from splitting of the MZMs, are localized near the end of the wire, and therefore couple strongly to the leads resulting in larger broadening compared to the sub-gap states. As expected, these extra states do not change the zero-bias conductance, which is controlled by the topological invariant.

	
\subsection{p-wave superconductors}
\label{pwave}

We will now consider the main focus of this paper, i.e., TR-invariant topological superconductors~\cite{Dumitrescu13} that can be realized by spin-triplet pairing, exhibiting equal spin pairing (ESP) $p$-wave superconductivity. These properties are conjectured to be present in the quasi-1D transition metal oxide Li$_{0.9}$Mo$_6$O$_{17}$, and some organic superconductors~\cite{limo1,triplet1,triplet2,Lebed_00,Lebed_13,Shinagawa_2007}. The hopping integrals along the crystallographic directions of these materials vary as, $t_x \gg t_y \gg t_z$, making them quasi-1D conductors. 

 The triplet ($S=1$) superconductivity of the p-wave wire can be represented by a matrix pair potential $\Delta_{\alpha\beta}(k)=\Delta\left[\textbf{d}(k)\cdot \bm{\sigma} \right ] _{\alpha\beta}$, where
$k=k_x$ is the 1D crystal-momentum.
The nature of the triplet component characterized by the vector $\mathbf d$ is odd (i.e. $\textbf{d}(k)=-\textbf{d}(-k)$), while $\Delta$ represents the magnitude.
Here, we choose $\mathbf d(k) =d(  k)\left(0,0,1\right )$, i.e., along
the $z$-direction in the spin space. The superconducting term in real space is then of the form: $ \mathrm{i}\,\Delta \left ( c_{ \ell+1,\uparrow}^{\dagger}\, c_{ \ell,\uparrow}^{\dagger}\,+\,c_{\ell+1,\downarrow}^{\dagger}\,
c_{ \ell,\downarrow}^{\dagger} \right )+ {\rm h.c.}$
This choice of the $\bm d$-vector represents a TR-invariant superconductor containing the ESP spin-triplet $p$-wave state.

The bulk Hamiltonian for a p-wave superconducting chain, with the order parameter described
in the previous paragraph, can be written in the Nambu basis (defined with the spinor $\Psi_{k}=(c_{k,\uparrow},c_{k,\downarrow},c_{-k,\downarrow}^{\dagger},-c_{-k,\uparrow}^{\dagger})^{T}$) as:
\begin{align}
\label{eq:H1DK}
&	{\cal{H}}^{1D}_{k}(\mu,\Delta, \mathbf V,\alpha_R) \nonumber \\
& = \left [\, \epsilon(k)-\mu+2\,t_x\,   \right ]\sigma_{0}\tau_z + \Delta\, \sin k \,\sigma_z\, \tau_{x}
 + {\cal{H}}^{Z}\,+ {\cal{H}}^{SO}\,.
\end{align}
Here, $\epsilon(k)=-2\,t_x \cos k $ is the single-particle kinetic energy,
and $ \Delta\,\sin k$ is the $p$-wave superconducting order parameter. 
In addition to p-wave superconductivity, the above Hamiltonian allows us to consider the 
effect of an electric-field-induced inversion-symmetry breaking spin-orbit interaction (SOI) term aligned in an arbitrary direction ${\mathbf{a}}$ in the spin space. This is written as 
\begin{align}{\cal{H}}^{SO}=\alpha_R \,\sin k
\left ({\mathbf{a}} \cdot \bm{\sigma}\right ) \tau_z\,.
\end{align}	
The Hamiltonian also allows us to consider the breaking of TR symmetry ,
through the Zeeman effect of an applied magnetic field, that is 
captured by the term
\begin{align}	
{\cal{H}}^{Z}= \left( \bm{V} \cdot \bm{\sigma} \right ) \tau_0 \,.
\end{align}	

The single-chain p-wave Hamiltonian described above can be generalized 
to model a quasi-1D system that is more relevant to the experimental materials (such as Li$_{0.9}$Mo$_6$O$_{17}$~\cite{Lebed_13,limo1}), by coupling 
multiple copies of Eq.~\ref{eq:H1DK} into a multichannel Hamiltonian, that is written as:
\begin{align}
  H^{Q1D}_{k;jj'} = & \,{\cal{H}}^{1\text{D}}_{k,j} \delta_{j, j'}+{\cal{H}}^{y}_{j,j'} \,,
\nonumber \\
{\cal{H}}^{1\text{D}}_{k,j} 
= & \, {\cal{H}}^{1D}_{k}
\left (\mu_j,\Delta_j,\mathbf V^j,\alpha_{R}^j \right ).
\label{eq:HQ1D}
\end{align}
The different chains in this quasi-1D bundle are coupled together 
with an amplitude $t_y$, given by the y-directional Hamiltonian 
\begin{align}
 {\cal{H}}^{y}_{j,j'}  
= &- t_y \, \tau_z   \left (\delta_{j,j'+1}+\delta_{j,j'-1} \right )
 \nonumber \\ & \quad 
+ \mathrm{i}\,\alpha_R'\, \sigma_y \, 
\tau_z \left (\delta_{j,j'+1}-\delta_{j,j'-1} \right ),
\end{align}
where $\alpha_R'$ represents the magnitude of the inter-chain Rashba SOI.
As we saw in the last subsection, coupling identical chains leads to an artificial decoupling
of sub-bands, that can lead to non-generic results. For this reason, we have introduced 
a $j$-dependence of the parameters of the single chains $\mathcal{O}_j=\mu_j,\Delta_j,\mathbf V^j,\alpha_{R}^j$, which is assumed to be of the form
\begin{align}
&\mathcal{O}_j=\bar{\mathcal{O}}\left (1-\tilde{j}\, \gamma \right ) ,
\end{align}
where $\bar{\mathcal{O}}$ is the average value of the parameters, $\gamma=0.1$, and $\tilde{j} = j-2$.

Combining all the ingredients discussed in this subsection so far, the total Hamiltonian in Eq.~\ref{eq:HQ1D} can be explicitly written out in the position space as:
\begin{widetext}
\begin{equation}
\label{nanoham2}
\begin{split}
H =  
&\sum_{\ell=1}^{N_x}
\sum_{j=1}^{N_y} 
{\psi}^{\dagger}_{\ell,j}
\left [  \left \lbrace 2(t_x+t_y) -\mu_{j} \right \rbrace \tau_{z} 
+ \mathbf{V}^j\cdot \bm{\sigma} \right ] {\psi}_{\ell,j}
- \sum_{\ell=1}^{N_x-1} \sum_{j=1}^{N_y}
\left [ {\psi}^{\dagger}_{\ell+1, j}
\left \lbrace t_{x}\, \tau_{z}\,
+ \frac{ \mathrm{i}\, \Delta_j}{2}\, \sigma_z\, \tau_x + 
\frac{ \mathrm{i}\, \alpha_{R}^{j}}{2}
\left (\mathbf{a}\cdot \bm{\sigma}  \right ) \tau_z \right \rbrace 
{\psi}_{\ell,j}+{\rm h.c.}  \right ]
\nonumber \\ &- \sum_{\ell=1}^{N_x} \sum_{j=1}^{N_y-1}
\left \{ {\psi}^{\dagger}_{\ell, j+1}
\left  ( \,t_{y}\,\tau_{z}\, -\mathrm{i}\,\alpha_R'\,\sigma_y\,\tau_z
\right  ){\psi}_{\ell,j} +{\rm h.c.} \right \} .
\end{split}
\end{equation}
\end{widetext}

The above Hamiltonian, in addition to obeying particle-hole symmetry $\mathcal{P}=\sigma_y\tau_y K$ (and taking $k\rightarrow -k$) that applies to any superconductor, obeys 
a TR symmetry for $\bm{V}=0$. The corresponding TR operator is $\mathcal{T}= \mathrm{i}\,\sigma_y\, K$ (plus performing the operation $k\rightarrow -k$), $K$ here denotes complex conjugation. In addition, depending on the presence or absence of spin-conservation (the latter caused by SOI), or inversion symmetry, the system can possess various 
mirror or chiral symmetries. This makes the p-wave system particularly interesting because 
it allows, in principle, turning the various symmetries on or off. by applying electric and magnetic fields, while all the while remaining in the topological phase.

We now study the signatures of the symmetry breaking phenomena in this topological p-wave 
superconductor in the subsubsections below. Specifically, we will compute the results for the spectrum, topological invariant, and conductance, similar to what we reviewed in the more familiar and simpler case of the Rashba nanowire in the previous subsection (illustrated in Fig.~\ref{1dnano}). 
For our numerics, we will choose the superconducting pairing amplitude to be $\Delta=0.5$ (in units where $t_x=1$), in addition to other parameters specified in the relevant subsubsection.

\subsubsection{Time-reversal and chiral symmetric case: $\mathbf V =\alpha_R'=0$}

The p-wave system of Eq.~\ref{eq:HQ1D}, with the restriction $\bm{V} =\alpha_R' = 0 $, is in the BDI symmetry class, similar to the Rashba nanowire studied in Sec.~\ref{swave}. This has a chiral symmetry operator $ \mathcal{S} = \sigma_z\, \tau_y$, which is different though, compared to the Rashba nanowire case. 

\begin{figure}[htb]
	\includegraphics[width = 0.4 \textwidth]{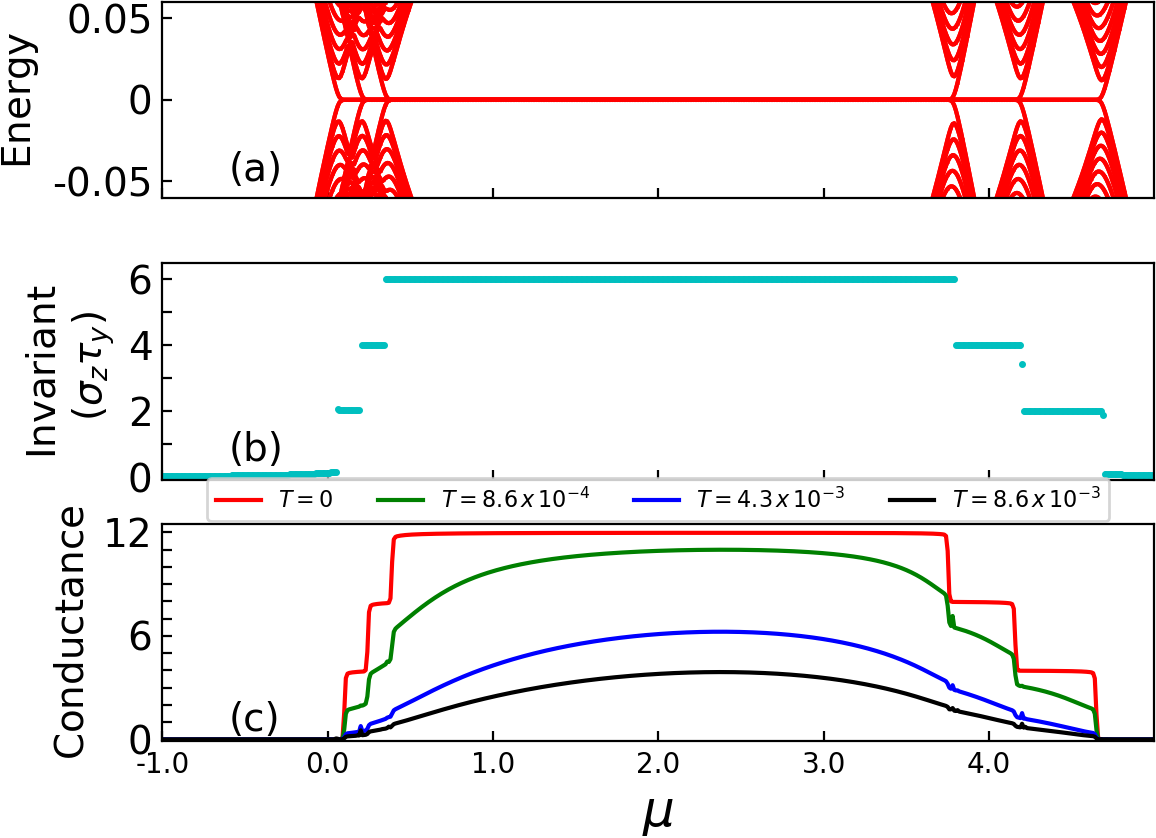}
\caption{\label{F2}
For quasi-1D p-wave superconductors of Eq.~\ref{eq:HQ1D}, described by the parameters $ t_y=0.1,\,\alpha_R =0.25, \, \alpha_R'=0.0,\,\mathbf V =0$, and ${\mathbf{a}}\,=\,(0,0,1)$, the figure shows the correspondence between the spectrum, the BDI invariant (associated with the chiral symmetry operator $\sigma_z\,\tau_{y}$), and the
zero-bias differential conductance, all of which have been plotted as functions of the   chemical potential $\mu$. All the temperature values are in units of $t_x$.}
\end{figure}

The spectrum of the model as a function of $\mu$ is shown in Fig.~\ref{F2}(a). This illustrates a sequence of bulk-gap closings, together with a set of zero-energy states, 
similar to the Rashba nanowire case of Sec.~\ref{swave}. However, in this case, we do 
not see significant sub-gap ABSs associated with the ends. The evolution 
of the spectrum can again be understood in terms of changing of the filling of sub-bands.  
Unlike the Rashba nanowire case, both spin components of a sub-band with p-wave 
filling are topological, with the same sign of the topological invariant. This can be understood from the chiral topological invariant, which is calculated analogous to the Rashba nanowire case, and is plotted in Fig.~\ref{F2}(b). From this plot, we see that the topological invariant jumps from zero to two, as $\mu$ increases from $\mu\sim 0$. This regime corresponds to both spin components of the lowest sub-band starting to fill. The fact that the topological invariant increases from zero to two indicates that both spin components of the sub-band contribute the same value to the topological invariant, which is different from the case of the Rashba nanowire. 
The difference can be understood in the simpler case with $\alpha_R=0$, where 
the Hamiltonian and the chiral symmetry $\mathcal{S}$ commute with the mirror symmetry operator $M = \mathrm{i} \left (\hat{\mathbf d}
\cdot \boldmath{\sigma} \right )\tau_0=\sigma_z\tau_0$. $M$ can thus be used to define a mirror-invariant as well~\cite{tudor15}. Thus a chiral topological invariant can be computed for each $\sigma_z=\pm 1$. Furthermore, the two sectors $\sigma_z=\pm 1$ can be mapped into each other by $\sigma_x$, which commutes with $\mathcal{S}$. This explains why the two sectors have the same topological invariant. The inclusion of non-zero $\alpha_R$ does not break the chiral symmetry (though it breaks the mirror symmetry), and therefore it cannot change the topological invariant. In fact, numerical results for $\alpha_R=0$ (not included here) appear qualitatively identical to Fig.~\ref{F2}.
The topological invariant begins to decreases at $\mu \gtrsim 4$, as the Fermi level crosses the tops of the bands.

The topological invariant indicates the number of MZMs that appear at each end 
of the system, which appear as zero-energy states in the spectrum shown in Fig.~\ref{F2}(a). 
The even-parity of the topological invariant can be understood to be a consequence of 
the TR symmetry, which constrains the MZMs to appear in Kramers pairs. The 
even number of MZMs appear as zero-bias conductance plotted in Fig.~\ref{F2}(c). Similar 
to the case of the Rashba nanowire, the zero-bias conductance closely tracks the 
number of MZMs, and provides a measurable indication of the change of the 
topological invariant (that cannot be measured directly). As in the case of the Rashba nanowire, the temperature dependence of the zero-bias conductance provides a sense of the energy scale with which the MZMs couple to the leads. As before, we find that the higher conductance peaks are more sensitive to temperature.

\begin{figure}[htb]
{\includegraphics[width = 0.4 \textwidth]{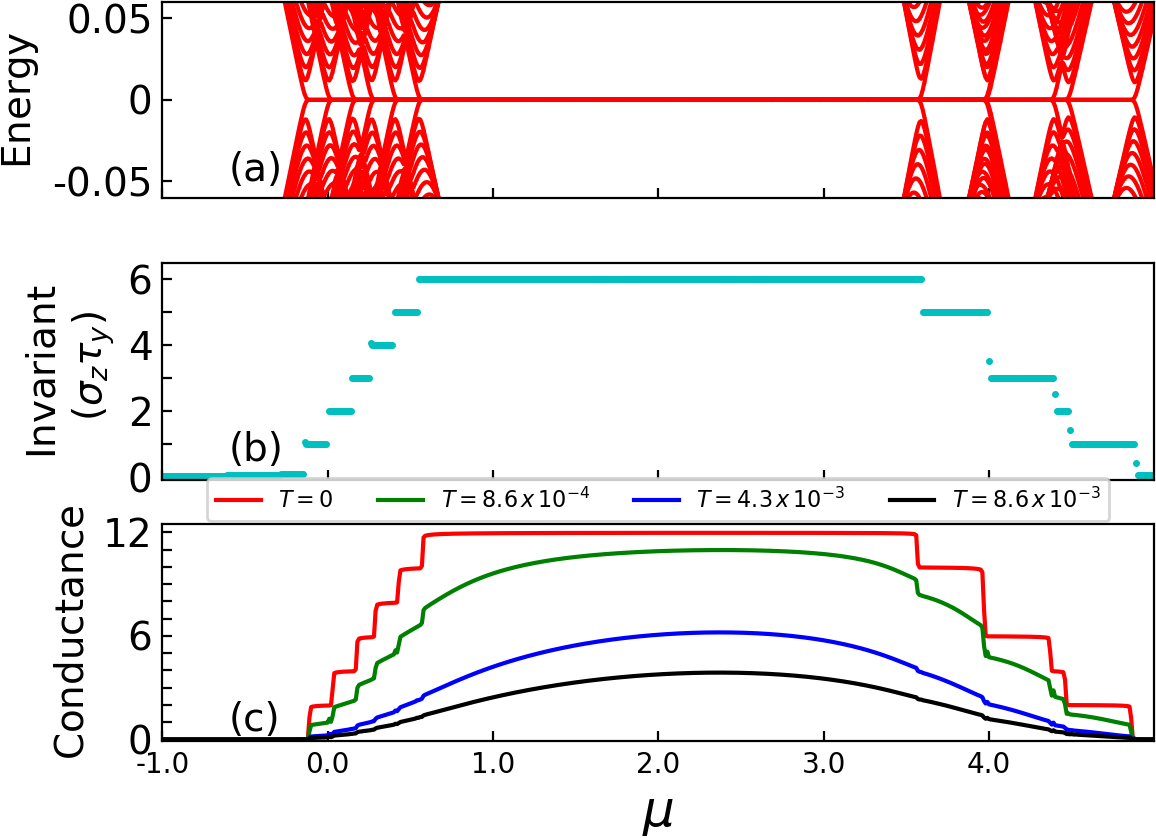}}
\caption{\label{F3}For quasi-1D p-wave superconductors of Eq.~\ref{eq:HQ1D}, described by the parameters $ t_y=0.1,\,\alpha_R =0.25, \, \alpha_R'=0$, ${\mathbf{a}}=(0,0,1)$, and $\mathbf V= (0,0.2,0)$, the figure shows the correspondence between the spectrum, the BDI invariant (associated with the chiral symmetry operator $\sigma_{z}\,\tau_{y}$), and the zero-bias differential conductance, all of which have been plotted as functions of the chemical potential $\mu$. All the temperature values are in units of $t_x$.}
\end{figure}

\subsubsection{Time-reversal broken chiral symmetric case:
$|{\mathbf{a}} \times {\hat{z}}|=\alpha_R'=0$}

The TR symmetry of the system, discussed in the previous sub-subsection,
can be broken by applying a Zeeman field $\mathbf V= (0,0.2,0)$.
The nanowire, however, still has chiral symmetry, encoded by the operator $\mathcal{S}=\sigma_z\tau_y$. The breaking of the TR symmetry splits the Kramer's degeneracy of the bulk states near the phase transitions seen in Fig.~\ref{F3}, such that the three phase transitions for $\mu<1$ (seen in Fig.~\ref{F2}(a)) are now split into six phase transitions
(see Fig.~\ref{F3}(a)). For the parameters chosen, the splitting of some of the higher chemical potential ($\mu>3$) transitions are too small to resolve. As seen 
in Fig.~\ref{F3}(b), these split-transitions are indeed topological phase transitions, as 
they are accompanied by a change of the topological invariant. The topological 
invariant, which is identical to the one calculated in the last sub-subsection, 
shows integer jumps for $\mu<1$, as opposed to the double jumps seen in the  
TR symmetric case discussed in the last sub-subsection.   

The MZMs that appear at the end because of the non-trivial topological invariant are 
no longer required to be Kramers degenerate. This allows both even and odd 
number of MZMs. This is apparent from the numerical result for the zero bias 
conductance shown in Fig.~\ref{F3}(c), where we see that many of the conductance 
steps that are multiples of $4e^2/h$ split into steps with height $2e^2/h$ 
associated with individual MZMs. Unfortunately, many of the smallest steps are washed out 
at the lowest temperature of $T\sim 10\,K$.

\begin{figure}[]
\includegraphics[width = 0.4 \textwidth]{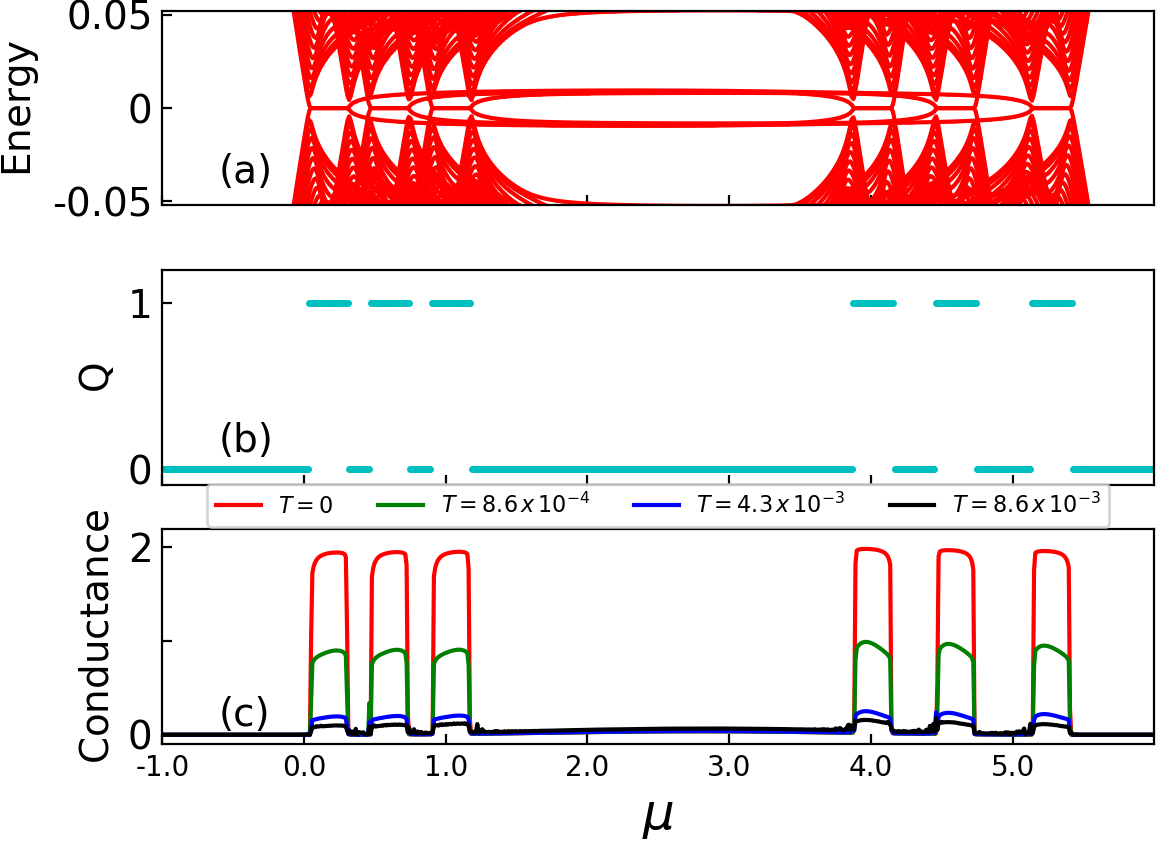}
\caption{\label{F4} For quasi-1D p-wave superconductors of Eq.~\ref{eq:HQ1D}, described by the parameters $t_y=0.3, \, \alpha_R=0.25$, $ \alpha_R^\prime=0 $, ${\mathbf{a}}\,=\,(1,1,0)$, and $\mathbf V= (0,0.1,0.1)$, the figure shows the correspondence between the spectrum, the Pfaffian invariant (as the system belongs to class D), and the zero-bias differential conductance, all of which have been plotted as functions of the chemical potential $\mu$.
All the temperature values are in units of $t_x$.}
\end{figure}

\subsubsection{Superconductors without symmetry}
\label{classD}

Changing the electric field symmetry of the system such that either the intra-chain 
Rashba coupling picks up a non-zero component (causing $|\mathbf {a}\times \hat{z}|\neq 0$), and/or generates an inter-chain $\alpha_R'\neq 0$, breaks all the chiral symmetries of the system. This places the system in the symmetry class D~\cite{kitaev-classi,*Schnyder}, which is the minimal symmetry for a superconductor, showing only particle-hole symmetry.  As seen from the spectrum plotted in Fig.~\ref{F4}(a), this leads 
to a structure of bulk-gap closings that is similar to Fig.~\ref{F3}(a), in the sense 
of showing six phase transitions at low chemical potential values ($\mu<2$). In this case, we also find six phase transitions above $\mu>2$.

However, unlike Fig.~\ref{F3}(a), the zero-energy states in Fig.~\ref{F3}(a) both appear or disappear at the subsequent phase transitions. This is consistent  with the fact 
that the topological invariant for class D is in the group $Z_2$, and therefore it takes only two values given by \cite{Kitaev_2001} $Q =\frac{1-\nu}{2}$,
where 
\begin{align}
& \nu = \text{sgn} \left[ Pf(Q_1) \,Pf(Q_2) \right ],\nonumber \\
& Q_1 = \mathcal{H}^{Q1D}(k)  \Big \vert_{k=0} \sigma_y\, \tau_y \,,\nonumber \\
& Q_2 = \mathcal{H}^{Q1D}(k) \Big \vert_{k=\pi}
\sigma_y\, \tau_y \,.
\end{align}
Here, $Pf(A)$ represents the Pfaffian of the matrix $A$ \cite{pf}. 
The resulting topological invariant $Q$ is plotted in Fig.~\ref{F4}(b), which shows 
that the system alternates between a topological phase (with one MZM at each end), and 
a trivial phase. Most of the trivial phases, seen in Fig.~\ref{F4}(b), correspond to 
the chemical potential range of Fig.~\ref{F4}(b) with an even value of the chiral topological invariant. These regions in the spectrum of Fig.~\ref{F4}(a) contain sub-gap states, that result from splitting the even number of MZMs by the breaking of chiral symmetry with SOI.

As in the earlier cases, the plot of the zero-bias conductance shown in Fig.~\ref{F4}(c)
indicates the topological invariant. The topological region with invariant $Q=1$ 
shows a conductance of $2e^2/h$, corresponding to a single MZM at each end of the wire.
The zero-bias conductance in this case shows a stronger suppression with temperature compared to its chiral symmetric counterpart in Fig.~\ref{F3}(c). This indicates that the residual MZM is the one that couples weakest to the lead.

\begin{figure}[]
\includegraphics[width = 0.4 \textwidth]{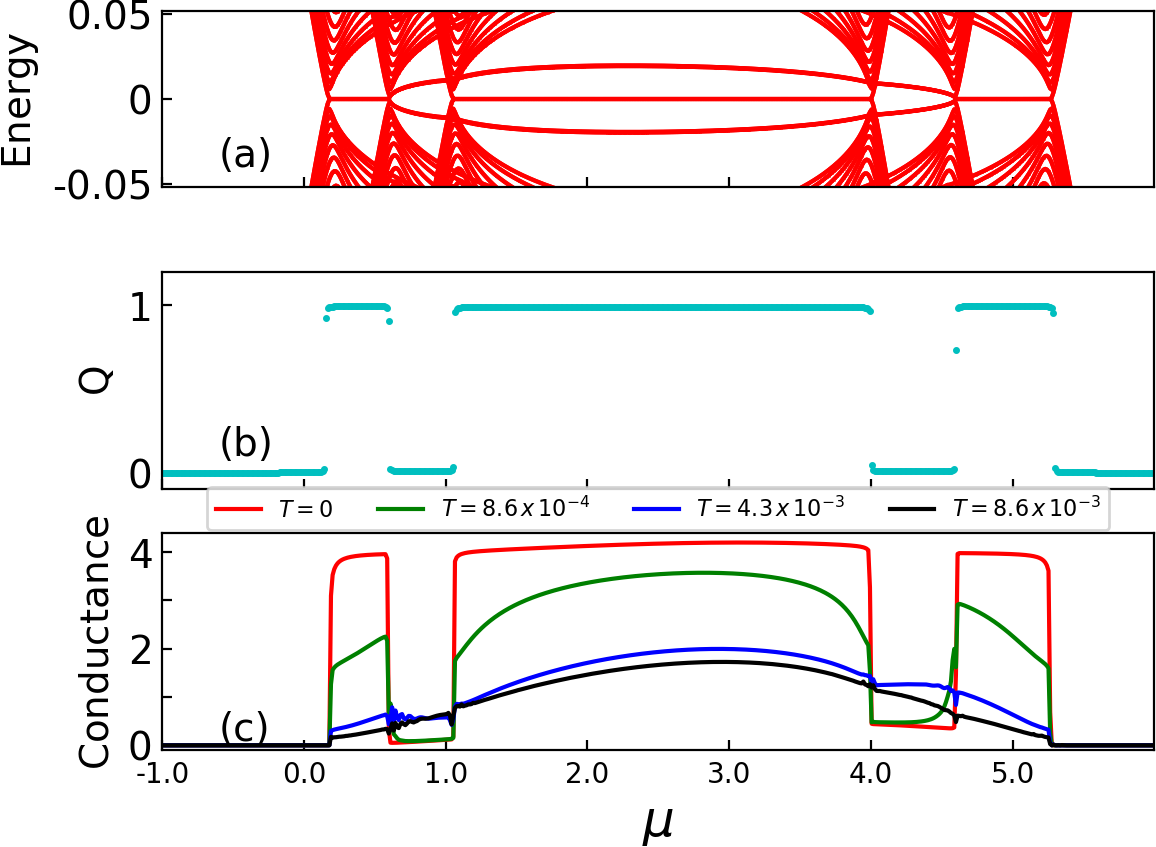}
\caption{\label{F5}For the quasi-1D p-wave superconductors of Eq.~\ref{eq:HQ1D}, described by the parameters $t_y=0.3, \alpha_R=0.25$, $ \alpha_R^\prime=0.1 $, ${\mathbf{a}}\,=\,(1,1,0)$, and $\mathbf{V}= 0$, the figure shows the correspondence between the spectrum, the DIII invariant ($Q$), and the zero-bias differential conductance, all of which have been plotted as functions of the chemical potential $\mu$.
All the temperature values are in units of $t_x$.}
\end{figure}


\subsubsection{Time-reversal preserving case}
\label{sectr}

Applying an electric field to generate an inter-chain Rashba SOI ($\alpha_R'\neq 0$),
as well as an intra-chain SOI ($\alpha_R\neq 0$), in the absence of a Zeeman field,
breaks all chiral and spin symmetries without breaking TR symmetry.
This leads to a superconductor in the symmetry class DIII. The symmetry class 
of the system becomes apparent from the spectrum shown in 
Fig.~\ref{F5}(a), which exhibits three phase transitions for $\mu\lesssim 1$, similar to 
the spectrum of the TR symmetric superconductor shown in 
Fig.~\ref{F3}(a). However, the chiral symmetry present in the case of Fig.~\ref{F3}(a), is broken here by the applied electric field. As a result, 
the MZMs disappear at alternate phase transition points, 
similar to the spectrum of the class D nanowires shown in Fig.~\ref{F4}(a).
Similar to the case of Sec.~\ref{classD}, the alternating presence of 
MZMs can be understood from the topological invariant for this case, which can take
two values (see Fig.~\ref{F4}(b)). As in the other cases, the topological 
invariant only changes at the phase transitions.
The topological invariant for the symmetry class DIII, shown in Fig.~\ref{F5}(b), is calculated as $Q=\frac{1-\nu}{2}$~\cite{DIII}, where
\begin{equation}
\nu = \det \left({\mathcal{U}}^K \right )\,\frac{Pf \left(\hat{\theta} (0) \right )}
{ Pf \left (\hat{\theta} (\pi) \right )}\,.
\end{equation} 
Here, $\hat{\theta} (0)$ and $\hat{\theta} (\pi)$ represent the matrix elements of 
the TR operator $\mathcal{T}$, in the basis of the occupied states, 
at $k=0$ and $k = \pi$, respectively. The matrix ${\mathcal{U}}^K $ in this basis 
is given by the so-called Kato propagator~\cite{DIII}: 
\begin{align}
{\mathcal{U}}^K(0,\pi) = \lim_{n\rightarrow \infty} \,\prod 
\limits_{\lambda=0}^{n} \, \mathcal{P}_{o}(k_\lambda)\,, 
\end{align}
where $\mathcal{P}_{o}(k_\lambda)$ is the projector into the occupied bands (negative energy) and $k_\lambda = \frac{ \lambda \, \pi}{n}$. 

The zero-bias conductance as a function 
of $\mu$, which is plotted in Fig.~\ref{F5}(c), provides a measurable indication 
of the topological invariant. Similar to the previous sub-subsection, this correspondence 
is apparent from the fact that the zero-bias conductance takes two values in the 
tunneling limit, corresponding to the two values of the topological invariant. 
The conductance in the topologically non-trivial regime takes a value $4e^2/h$ 
corresponding to two MZMs at each end. The doubled value of the quantized 
conductance, in contrast with the case in Sec.~\ref{classD}, is a 
result of the Kramers degeneracy associated with TR symmetry.
Relative to the spectrum of the multiple Kramers pairs of MZMs shown in 
Fig.~\ref{F2}(a) (for the chiral and TR symmetric p-wave superconductor),
Kramers pairs of ABSs or MZMs can be seen in the spectrum in Fig.~\ref{F5}(a).   
Thus the DIII class topological invariant and the structure of ABSs in Fig.~\ref{F5} can be understood from the chiral symmetric topological invariant at the corresponding 
values of chemical potential in Fig.~\ref{F2}(b).

\section{Signatures of multiple MZMs in Andreev spectroscopy}
\label{cavity}

In this section, we focus on the effect of the symmetry-breaking on the ABS spectra of the JJs of superconducting nanowires.
A JJ is created by introducing a weak link with tunneling amplitude 
$\gamma$, which has negligible conductance compared to the rest of the superconducting wire.
We can therefore assume that the supercurrent has a negligible contribution to the superconducting phase drop around the wire, which can be controlled by a flux-loop~\cite{felix}. 
The ABS spectrum of a JJ generates features in the ac absorption, that can be measured by several techniques such as microwave-spectroscopy~\cite{van_Woerkom} and two-tone spectroscopy~\cite{urbina,Devoret}.

Following the argument in the last paragraph, the superconducting phase 
difference $\phi$ across the JJ, generated by the magnetic flux-loops, is introduced through a modified Hamiltonian corresponding to Eq.~\ref{eq:HQ1D}:
\begin{widetext}	
 \begin{align}
\label{nanoham3}
	H = & \sum_{\ell=1}^{N_x}\sum_{j=1}^{N_y} {\psi}^{\dagger}_{\ell,j}\left ((-\mu+2(t_x+t_y)) \,\tau_{z}+\Delta_s\, \tau_{x}^{\ell}+V_{x} \,\sigma_{x} \right ){\psi}_{\ell,j}
- \sum_{\ell=1}^{N_x-1} \sum_{j=1}^{N_y}
	\left \{ {\psi}^{\dagger}_{\ell+1, j}
\left (t_{x}+  \mathrm{i}\,\alpha\, \sigma_{y}\right )\tau_{z}\, {\psi}_{\ell,j}+{\rm h.c.}  \right\}
	\nonumber \\	&
	- \sum_{\ell=1}^{N_x} \sum_{j=1}^{N_y-1}
\left ( {\psi}^{\dagger}_{\ell, j+1} \,t_{y}\,\tau_{z}\, {\psi}_{\ell,j} +{\rm h.c.} \right ) - \gamma \sum_{j=1}^{N_y}
\left ( {\psi}^{\dagger}_{N_x, j} \,\tau_{z}\, {\psi}_{1,j}+{\rm h.c.} \right),
\end{align}
\end{widetext}	
where $\tau_x^{\ell} =\begin{pmatrix} 0 & 
e^{\mathrm{i}\, \ell\,\phi/N_x} 
\\ e^{-\mathrm{i}\, \ell\,\phi/N_x} 
& 0\end{pmatrix}$. Furthermore, the phase difference $\phi$ across the JJ is controlled by the magnetic flux $\Phi$ in the superconducting loop through the relation
$\phi=2\,\pi\,\Phi/\Phi_0$, where $\Phi_0= \frac{h\,c} {2\,e}$ is the superconducting flux quantum. 
The $\ell$ dependence of the matrix $\tau_x^{\ell}$ can be understood as a winding of the superconducting phase around the loop, which is needed to ensure that the superconducting phase difference between the ends $\ell=1$ and $\ell=N_x$ of the JJ is equal to $\Phi$.

\begin{figure}[]
\includegraphics[width =0.4 \textwidth]{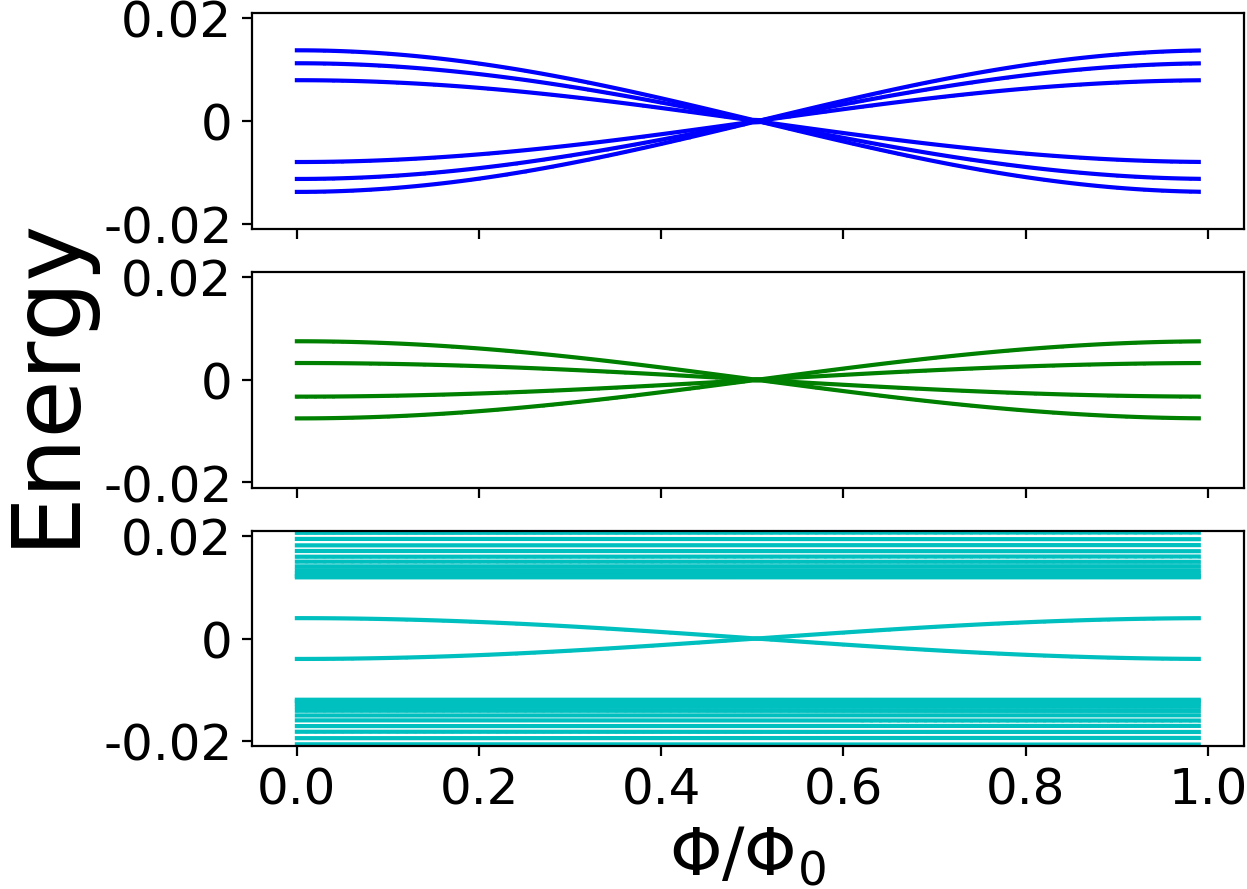}
\caption{\label{sring}
ABS spectrum of a quasi-1D Rashba nanowire, with the same parameters as in Fig.~\ref{1dnano}, as a function of the flux across a Josephson junction. The ends of the multi-chain ring have a weak coupling strength of $\gamma=0.1$.
The three panels have different values of the chemical potential: 
$\mu=0.1$ (top), $\mu= -0.52 $ (middle), and $\mu = 0.4$ (bottom). The spectrum shows degeneracies of $6$, $4$, and $2$, at flux $\Phi=\Phi_0/2$. These correspond to $3$, $2$, and $1$ MZM(s) at each end of the nanowire in the open boundary case, as seen in Fig.~\ref{1dnano}.}
\end{figure}

The ABS spectrum of the above Hamiltonian is plotted in Fig.~\ref{sring} for 
several chemical potential values, and it shows zero-energy crossings at
$\frac{\phi}{2\,\pi} = 1/2$. The class BDI topological invariant for 
the bulk Hamiltonian, with the same parameters as used to plot the spectrum in Fig.~\ref{1dnano}, shows clearly that the number of zero-energy ABSs at
$\frac{\phi}{2\,\pi}=1/2$, for the different chemical potentials, is twice the topological invariant. This can be understood by noting that, prior to introducing the tunnel coupling (i.e., at $\gamma=0$), the topological invariant in the BDI symmetry class equals the number of MZMs at each end of the JJ. Introducing the tunnel coupling $\gamma$ across the JJ splits the pairs of MZMs into ABSs with energies that are typically non-zero, except at $\phi=\pi$.

We can understand the above argument more explicitly by applying a gauge transformation, $U=e^{ \frac{\mathrm{i}\,\ell \,\phi\,\tau_z}{2\,N_x}}$, to each site $\ell$ of the lattice. This eliminates the $\ell$ dependence of the SC phase in $\tau_x^\ell$, in favour of introducing a phase $e^{\mathrm{i}\,\phi\,\tau_z/2}$ into the tunneling 
term proportional to $\gamma$ (in Eq.~\ref{nanoham3}). Then this term becomes 
proportional to $\mathrm{i}\,\gamma$ for phase $\phi=\pi$. Since this term commutes with the 
chirality operator $\mathcal{S}=\sigma_y\,\tau_y$ (see Sec.~\ref{swave}), and
the MZMs for the BDI class are eigenstates of $\mathcal{S}$ with eigenvalues $\mathcal{S}=\pm 1$ in the absence of tunneling (i.e., $\gamma=0$), these MZMs cannot be hybridized by the coupling $\gamma$ at phase $\phi=\pi$. As a result, they appear as zero-energy ABSs, as shown in Fig.~\ref{sring}.

\begin{figure}[]
\includegraphics[width =0.4\textwidth]{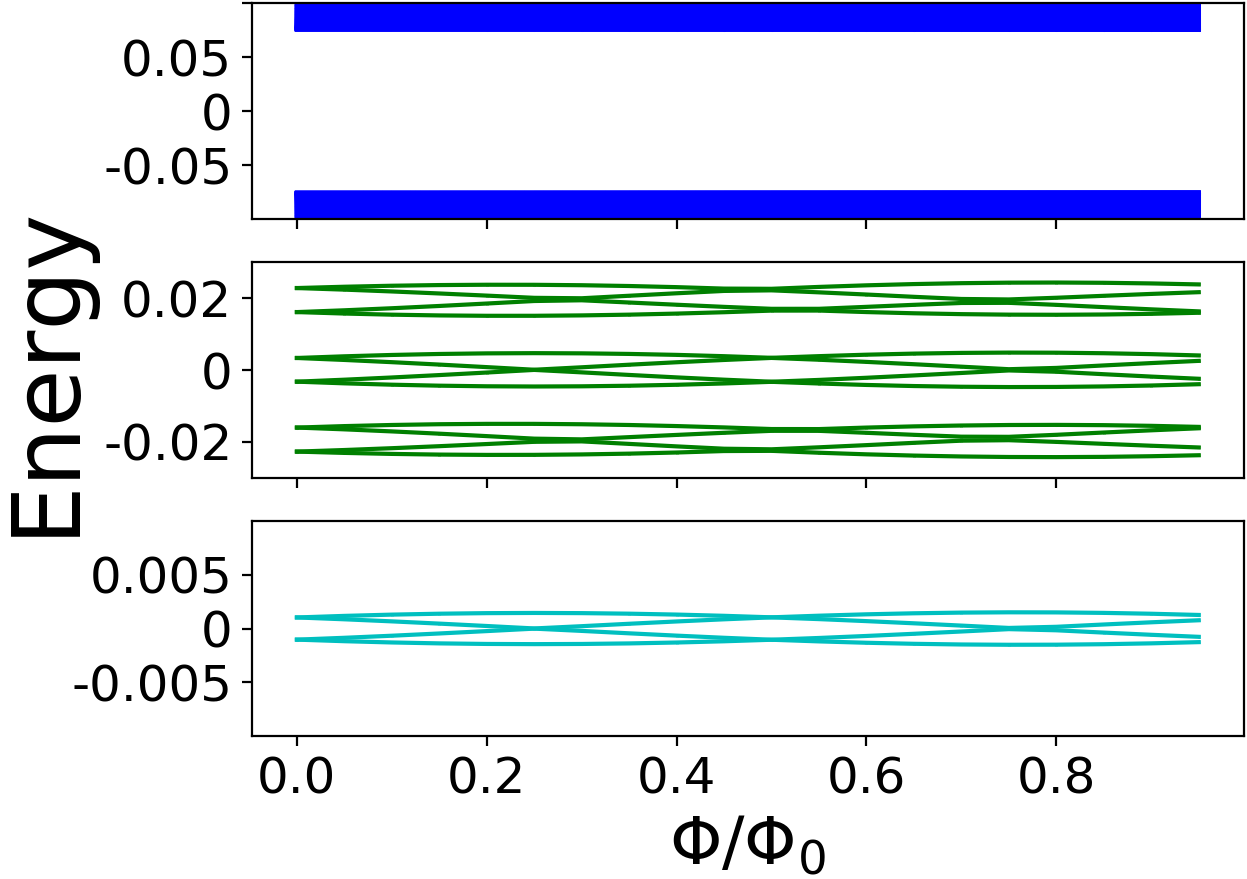}
\caption{\label{pring}
ABS spectrum of a quasi-1D p-wave superconductor, with the same parameters as in Fig.~\ref{F5}, as a function of the flux across a Josephson junction. The ends of the multi-chain ring has a weak coupling strength of $\gamma=0.1$.
The three panels have different values of the chemical potential: 
$\mu=0.0$ (top), $\mu= 2.0$ (middle), and $\mu =5.0$ (bottom). The spectrum shows degeneracies of $0$, $4$, and $4$, at flux $\Phi=\Phi_0/2$. These correspond to $0$, $2$, and $2$ MZMs at each end of the nanowire in the open boundary case, as seen in Fig.~\ref{F5}. The middle panel shows additional states due to the presence of ABSs (also seen in Fig.~\ref{F5}(a)), which also exist in the open boundary case at low energies.}
\end{figure}

We now analyze the p-wave system treated earlier in Sec.~\ref{pwave}. Introducing a JJ into the system with a phase difference $\phi$ (similar to Eq.~\ref{nanoham3}), leads to the modified Hamiltonian:
\begin{widetext}	
\begin{align}
\label{nanoham4}
H =  &\sum_{\ell=1}^{N_x}
	\sum \limits_{j=1}^{N_y} 
	{\psi}^{\dagger}_{\ell,j} \left[
	\left \lbrace (-\mu_j +2 \left (t_x+t_y \right ) \right \rbrace
\tau_{z} + \mathbf{V}^j\cdot\mathbf{\sigma} \right ] {\psi}_{\ell,j}
- \sum_{\ell=1}^{N_x-1} \sum_{j=1}^{N_y}
	\left [ {\psi}^{\dagger}_{\ell+1, j}\left \lbrace t_{x}\,\tau_{z}
+ \frac{\mathrm{i}\, \Delta_j}{2} \,\sigma_z \, \tau_x^{\ell} 
+ \frac{  \mathrm{i}\, \alpha_{R}^{j}}{2}\left (\mathbf{a}\cdot\mathbf{\sigma} \right )
\tau_z \right \rbrace {\psi}_{\ell,j}+{\rm h.c.}  \right ] \nonumber \\ 
&- \sum_{\ell=1}^{N_x} \sum_{j=1}^{N_y-1}
	\left \{ {\psi}^{\dagger}_{\ell, j+1} 
	\left( t_{y}\,\tau_{z}\, -\mathrm{i}\,\alpha_R'\,\sigma_y\tau_z \right)
	{\psi}_{\ell,j} +{\rm h.c.} \right \} - \gamma \sum_{j=1}^{N_y}
\left ( {\psi}^{\dagger}_{N_x, j} \,\tau_{z}\, {\psi}_{1,j}+{\rm h.c.} \right ) ,
	\end{align}
\end{widetext}	
with  $\tau_x^{\ell}$ having the same meaning as in Eq.~\ref{nanoham3}. 
We consider the same parameters as in Sec.~\ref{sectr}.
Similar to the case of the s-wave superconductor with Rashba SOI 
studied in Fig.~\ref{sring}, the tunnel coupling proportional to $\gamma$ splits 
the end MZMs into finite-energy ABSs, as seen in Fig.~\ref{pring}. However, unlike the s-wave case, we see the ABSs cross zero energy at two points: at $\Phi = \Phi_0/4$ and $3\,\Phi_0/4$. 
The number of ABSs merging at zero energy varies with the 
chemical potential, as shown in the three different panels of Fig.~\ref{pring}.
Comparison of the values of the chemical potential with the topological invariant in Fig.~\ref{F2}, shows clearly that the number of zero-energy ABSs at $\Phi=\Phi_0/4$ and $ 3\Phi_0/4$ is twice the topological invariant, similar to the case of the Rashba nanowire.
The middle panel also shows low energy ABSs, in addition to the zero-energy 
crossings associated with the MZMs.

The results shown in Figs.~\ref{sring} and \ref{pring} for the highly symmetric cases show that the Andreev spectra also show multiple zero-energy ABSs similar to the corresponding conductance plots. Therefore, we expect these results to be somewhat fragile in the sense that breaking the symmetries will lower the number of Andreev crossings (similar to what is seen for ZBCPs).

\section{Discussion and conclusion}
\label{end}

In this paper, we have studied the effect of symmetry-breaking field for 
multichannel p-wave superconductors that have a high degree of symmetry.
These symmetries allow the possibility of topological superconductivity 
with integer topological invariants with an integer number of MZMs. 
We find that the ZBCP reflects this topological invariant, as both vary  
synchronously with the chemical potential. Breaking the symmetries systematically, by applying strain and magnetic fields, leads to reducing the conductance to 
lower integer values by splitting some of the MZMs. The integer topological 
invariants also appear to manifest as zero-energy crossings of Andreev spectra for the highly symmetric topological superconductors. For the examples we consider, the number of 
zero-energy crossings of ABSs corresponds to the topological invariant, 
similar to the zero-bias conductance.  The crossings of the Andreev spectra in
topological JJs are expected to be measurable through 
recent advances in Andreev spectrocopy~\cite{van_Woerkom,tosi-urbina,dev}. The changes in the signatures (such as zero-bias conductance and Andreev spectroscopy) of the topological invariant can elevate the fingerprints of the MZMs to a rich structure, where the observables vary over several quantized values in a multi-dimensional phase space. 

While the signatures for integer topological invariants appear to be quite 
robust to variations in the Hamiltonian, the numerical examples we considered 
so far do not include disorder. How far these predictions hold up to realistic 
disorder in these systems will be an interesting direction for future work. In 
addition, it will be interesting to see if the zero-bias conductance with normal 
leads translates to a robust signature for superconducting leads, as with the 
case of non-degenerate Majorana modes~\cite{glazman-oppen}.
Explicit computation of the detection of the ABSs through a 
cavity-response experiment is also left for future work.

While realization of quasi-1D triplet superconductors~\cite{limo1,Lebed_13,Lebed_00,triplet2,Shinagawa_2007,
raghu} is in a preliminary state relative to semiconductor systems (which have seen rapid progress), the possibility of multiple MZMs with the chiral and time-reversal symmetry makes them a rich playground for controlling splitting of MZMs with external perturbations. Additionally, synthetic platforms have been proposed for realizing class DIII MZMs in semiconductor systems~\cite{PhysRevB.89.220504,ips-kramers}. We expect the transport signatures described in this paper will apply to these platforms as well.
	
\section{Acknowledgments}
	
We thank Pablo San-Jose for collaboration in the initial stages of the project. We also thank Denis Chevallier for useful discussions. Pfaffians were calculated using a code written by Bas Nijholt. ABR acknowledges the computational facilities provided by IIT-KGP, where parts of the work were carried out. JS acknowledges support from 	NSF
DMR1555135 (CAREER).

	\bibliography{biblio}

\begin{thebibliography}{61}%
\makeatletter
\providecommand \@ifxundefined [1]{%
 \@ifx{#1\undefined}
}%
\providecommand \@ifnum [1]{%
 \ifnum #1\expandafter \@firstoftwo
 \else \expandafter \@secondoftwo
 \fi
}%
\providecommand \@ifx [1]{%
 \ifx #1\expandafter \@firstoftwo
 \else \expandafter \@secondoftwo
 \fi
}%
\providecommand \natexlab [1]{#1}%
\providecommand \enquote  [1]{``#1''}%
\providecommand \bibnamefont  [1]{#1}%
\providecommand \bibfnamefont [1]{#1}%
\providecommand \citenamefont [1]{#1}%
\providecommand \href@noop [0]{\@secondoftwo}%
\providecommand \href [0]{\begingroup \@sanitize@url \@href}%
\providecommand \@href[1]{\@@startlink{#1}\@@href}%
\providecommand \@@href[1]{\endgroup#1\@@endlink}%
\providecommand \@sanitize@url [0]{\catcode `\\12\catcode `\$12\catcode
  `\&12\catcode `\#12\catcode `\^12\catcode `\_12\catcode `\%12\relax}%
\providecommand \@@startlink[1]{}%
\providecommand \@@endlink[0]{}%
\providecommand \url  [0]{\begingroup\@sanitize@url \@url }%
\providecommand \@url [1]{\endgroup\@href {#1}{\urlprefix }}%
\providecommand \urlprefix  [0]{URL }%
\providecommand \Eprint [0]{\href }%
\providecommand \doibase [0]{https://doi.org/}%
\providecommand \selectlanguage [0]{\@gobble}%
\providecommand \bibinfo  [0]{\@secondoftwo}%
\providecommand \bibfield  [0]{\@secondoftwo}%
\providecommand \translation [1]{[#1]}%
\providecommand \BibitemOpen [0]{}%
\providecommand \bibitemStop [0]{}%
\providecommand \bibitemNoStop [0]{.\EOS\space}%
\providecommand \EOS [0]{\spacefactor3000\relax}%
\providecommand \BibitemShut  [1]{\csname bibitem#1\endcsname}%
\let\auto@bib@innerbib\@empty
\bibitem [{\citenamefont {Alicea}(2012)}]{m1}%
  \BibitemOpen
  \bibfield  {author} {\bibinfo {author} {\bibfnamefont {J.}~\bibnamefont
  {Alicea}},\ }\bibfield  {title} {\bibinfo {title} {New directions in the
  pursuit of majorana fermions in solid state systems},\ }\href
  {https://doi.org/10.1088/0034-4885/75/7/076501} {\bibfield  {journal}
  {\bibinfo  {journal} {Reports on Progress in Physics}\ }\textbf {\bibinfo
  {volume} {75}},\ \bibinfo {pages} {076501} (\bibinfo {year}
  {2012})}\BibitemShut {NoStop}%
\bibitem [{\citenamefont {Elliott}\ and\ \citenamefont {Franz}(2015)}]{m2}%
  \BibitemOpen
  \bibfield  {author} {\bibinfo {author} {\bibfnamefont {S.~R.}\ \bibnamefont
  {Elliott}}\ and\ \bibinfo {author} {\bibfnamefont {M.}~\bibnamefont
  {Franz}},\ }\bibfield  {title} {\bibinfo {title} {Colloquium: Majorana
  fermions in nuclear, particle, and solid-state physics},\ }\href
  {https://doi.org/10.1103/RevModPhys.87.137} {\bibfield  {journal} {\bibinfo
  {journal} {Rev. Mod. Phys.}\ }\textbf {\bibinfo {volume} {87}},\ \bibinfo
  {pages} {137} (\bibinfo {year} {2015})}\BibitemShut {NoStop}%
\bibitem [{\citenamefont {Stanescu}\ and\ \citenamefont {Tewari}(2013)}]{m4}%
  \BibitemOpen
  \bibfield  {author} {\bibinfo {author} {\bibfnamefont {T.~D.}\ \bibnamefont
  {Stanescu}}\ and\ \bibinfo {author} {\bibfnamefont {S.}~\bibnamefont
  {Tewari}},\ }\bibfield  {title} {\bibinfo {title} {Majorana fermions in
  semiconductor nanowires: fundamentals, modeling, and experiment},\ }\href
  {https://doi.org/10.1088/0953-8984/25/23/233201} {\bibfield  {journal}
  {\bibinfo  {journal} {Journal of Physics: Condensed Matter}\ }\textbf
  {\bibinfo {volume} {25}},\ \bibinfo {pages} {233201} (\bibinfo {year}
  {2013})}\BibitemShut {NoStop}%
\bibitem [{\citenamefont {Leijnse}\ and\ \citenamefont {Flensberg}(2012)}]{m5}%
  \BibitemOpen
  \bibfield  {author} {\bibinfo {author} {\bibfnamefont {M.}~\bibnamefont
  {Leijnse}}\ and\ \bibinfo {author} {\bibfnamefont {K.}~\bibnamefont
  {Flensberg}},\ }\bibfield  {title} {\bibinfo {title} {Introduction to
  topological superconductivity and majorana fermions},\ }\href
  {https://doi.org/10.1088/0268-1242/27/12/124003} {\bibfield  {journal}
  {\bibinfo  {journal} {Semiconductor Science and Technology}\ }\textbf
  {\bibinfo {volume} {27}},\ \bibinfo {pages} {124003} (\bibinfo {year}
  {2012})}\BibitemShut {NoStop}%
\bibitem [{\citenamefont {Beenakker}(2013)}]{m6}%
  \BibitemOpen
  \bibfield  {author} {\bibinfo {author} {\bibfnamefont {C.}~\bibnamefont
  {Beenakker}},\ }\bibfield  {title} {\bibinfo {title} {Search for majorana
  fermions in superconductors},\ }\href
  {https://doi.org/10.1146/annurev-conmatphys-030212-184337} {\bibfield
  {journal} {\bibinfo  {journal} {Annual Review of Condensed Matter Physics}\
  }\textbf {\bibinfo {volume} {4}},\ \bibinfo {pages} {113} (\bibinfo {year}
  {2013})}\BibitemShut {NoStop}%
\bibitem [{\citenamefont {Nayak}\ \emph {et~al.}(2008)\citenamefont {Nayak},
  \citenamefont {Simon}, \citenamefont {Stern}, \citenamefont {Freedman},\ and\
  \citenamefont {Das~Sarma}}]{nayakrmp}%
  \BibitemOpen
  \bibfield  {author} {\bibinfo {author} {\bibfnamefont {C.}~\bibnamefont
  {Nayak}}, \bibinfo {author} {\bibfnamefont {S.~H.}\ \bibnamefont {Simon}},
  \bibinfo {author} {\bibfnamefont {A.}~\bibnamefont {Stern}}, \bibinfo
  {author} {\bibfnamefont {M.}~\bibnamefont {Freedman}},\ and\ \bibinfo
  {author} {\bibfnamefont {S.}~\bibnamefont {Das~Sarma}},\ }\bibfield  {title}
  {\bibinfo {title} {Non-abelian anyons and topological quantum computation},\
  }\href {https://doi.org/10.1103/RevModPhys.80.1083} {\bibfield  {journal}
  {\bibinfo  {journal} {Rev. Mod. Phys.}\ }\textbf {\bibinfo {volume} {80}},\
  \bibinfo {pages} {1083} (\bibinfo {year} {2008})}\BibitemShut {NoStop}%
\bibitem [{\citenamefont {Fu}\ and\ \citenamefont {Kane}(2008)}]{fu-kane}%
  \BibitemOpen
  \bibfield  {author} {\bibinfo {author} {\bibfnamefont {L.}~\bibnamefont
  {Fu}}\ and\ \bibinfo {author} {\bibfnamefont {C.~L.}\ \bibnamefont {Kane}},\
  }\bibfield  {title} {\bibinfo {title} {Superconducting proximity effect and
  majorana fermions at the surface of a topological insulator},\ }\href
  {https://doi.org/10.1103/PhysRevLett.100.096407} {\bibfield  {journal}
  {\bibinfo  {journal} {Phys. Rev. Lett.}\ }\textbf {\bibinfo {volume} {100}},\
  \bibinfo {pages} {096407} (\bibinfo {year} {2008})}\BibitemShut {NoStop}%
\bibitem [{\citenamefont {Sau}\ \emph {et~al.}(2010)\citenamefont {Sau},
  \citenamefont {Lutchyn}, \citenamefont {Tewari},\ and\ \citenamefont
  {Das~Sarma}}]{sau}%
  \BibitemOpen
  \bibfield  {author} {\bibinfo {author} {\bibfnamefont {J.~D.}\ \bibnamefont
  {Sau}}, \bibinfo {author} {\bibfnamefont {R.~M.}\ \bibnamefont {Lutchyn}},
  \bibinfo {author} {\bibfnamefont {S.}~\bibnamefont {Tewari}},\ and\ \bibinfo
  {author} {\bibfnamefont {S.}~\bibnamefont {Das~Sarma}},\ }\bibfield  {title}
  {\bibinfo {title} {Robustness of majorana fermions in proximity-induced
  superconductors},\ }\href {https://doi.org/10.1103/PhysRevB.82.094522}
  {\bibfield  {journal} {\bibinfo  {journal} {Phys. Rev. B}\ }\textbf {\bibinfo
  {volume} {82}},\ \bibinfo {pages} {094522} (\bibinfo {year}
  {2010})}\BibitemShut {NoStop}%
\bibitem [{\citenamefont {Lutchyn}\ \emph {et~al.}(2011)\citenamefont
  {Lutchyn}, \citenamefont {Stanescu},\ and\ \citenamefont
  {Das~Sarma}}]{lutchyn}%
  \BibitemOpen
  \bibfield  {author} {\bibinfo {author} {\bibfnamefont {R.~M.}\ \bibnamefont
  {Lutchyn}}, \bibinfo {author} {\bibfnamefont {T.~D.}\ \bibnamefont
  {Stanescu}},\ and\ \bibinfo {author} {\bibfnamefont {S.}~\bibnamefont
  {Das~Sarma}},\ }\bibfield  {title} {\bibinfo {title} {Search for majorana
  fermions in multiband semiconducting nanowires},\ }\href
  {https://doi.org/10.1103/PhysRevLett.106.127001} {\bibfield  {journal}
  {\bibinfo  {journal} {Phys. Rev. Lett.}\ }\textbf {\bibinfo {volume} {106}},\
  \bibinfo {pages} {127001} (\bibinfo {year} {2011})}\BibitemShut {NoStop}%
\bibitem [{\citenamefont {Oreg}\ \emph {et~al.}(2010)\citenamefont {Oreg},
  \citenamefont {Refael},\ and\ \citenamefont {von Oppen}}]{oreg}%
  \BibitemOpen
  \bibfield  {author} {\bibinfo {author} {\bibfnamefont {Y.}~\bibnamefont
  {Oreg}}, \bibinfo {author} {\bibfnamefont {G.}~\bibnamefont {Refael}},\ and\
  \bibinfo {author} {\bibfnamefont {F.}~\bibnamefont {von Oppen}},\ }\bibfield
  {title} {\bibinfo {title} {Helical liquids and majorana bound states in
  quantum wires},\ }\href {https://doi.org/10.1103/PhysRevLett.105.177002}
  {\bibfield  {journal} {\bibinfo  {journal} {Phys. Rev. Lett.}\ }\textbf
  {\bibinfo {volume} {105}},\ \bibinfo {pages} {177002} (\bibinfo {year}
  {2010})}\BibitemShut {NoStop}%
\bibitem [{\citenamefont {Kitaev}(2009)}]{kitaev-classi}%
  \BibitemOpen
  \bibfield  {author} {\bibinfo {author} {\bibfnamefont {A.}~\bibnamefont
  {Kitaev}},\ }\bibfield  {title} {\bibinfo {title} {Periodic table for
  topological insulators and superconductors},\ }\href
  {https://doi.org/10.1063/1.3149495} {\bibfield  {journal} {\bibinfo
  {journal} {AIP Conference Proceedings}\ }\textbf {\bibinfo {volume} {1134}},\
  \bibinfo {pages} {22} (\bibinfo {year} {2009})}\BibitemShut {NoStop}%
\bibitem [{\citenamefont {Schnyder}\ \emph {et~al.}(2009)\citenamefont
  {Schnyder}, \citenamefont {Ryu}, \citenamefont {Furusaki},\ and\
  \citenamefont {Ludwig}}]{Schnyder}%
  \BibitemOpen
  \bibfield  {author} {\bibinfo {author} {\bibfnamefont {A.~P.}\ \bibnamefont
  {Schnyder}}, \bibinfo {author} {\bibfnamefont {S.}~\bibnamefont {Ryu}},
  \bibinfo {author} {\bibfnamefont {A.}~\bibnamefont {Furusaki}},\ and\
  \bibinfo {author} {\bibfnamefont {A.~W.~W.}\ \bibnamefont {Ludwig}},\
  }\bibfield  {title} {\bibinfo {title} {Classification of topological
  insulators and superconductors},\ }\href {https://doi.org/10.1063/1.3149481}
  {\bibfield  {journal} {\bibinfo  {journal} {AIP Conference Proceedings}\
  }\textbf {\bibinfo {volume} {1134}},\ \bibinfo {pages} {10} (\bibinfo {year}
  {2009})}\BibitemShut {NoStop}%
\bibitem [{\citenamefont {Zhang}\ \emph {et~al.}(2013)\citenamefont {Zhang},
  \citenamefont {Kane},\ and\ \citenamefont {Mele}}]{kane-mele}%
  \BibitemOpen
  \bibfield  {author} {\bibinfo {author} {\bibfnamefont {F.}~\bibnamefont
  {Zhang}}, \bibinfo {author} {\bibfnamefont {C.~L.}\ \bibnamefont {Kane}},\
  and\ \bibinfo {author} {\bibfnamefont {E.~J.}\ \bibnamefont {Mele}},\
  }\bibfield  {title} {\bibinfo {title} {Time-reversal-invariant topological
  superconductivity and majorana kramers pairs},\ }\href
  {https://doi.org/10.1103/PhysRevLett.111.056402} {\bibfield  {journal}
  {\bibinfo  {journal} {Phys. Rev. Lett.}\ }\textbf {\bibinfo {volume} {111}},\
  \bibinfo {pages} {056402} (\bibinfo {year} {2013})}\BibitemShut {NoStop}%
\bibitem [{\citenamefont {Fidkowski}\ and\ \citenamefont
  {Kitaev}(2010)}]{fidkowski}%
  \BibitemOpen
  \bibfield  {author} {\bibinfo {author} {\bibfnamefont {L.}~\bibnamefont
  {Fidkowski}}\ and\ \bibinfo {author} {\bibfnamefont {A.}~\bibnamefont
  {Kitaev}},\ }\bibfield  {title} {\bibinfo {title} {Effects of interactions on
  the topological classification of free fermion systems},\ }\href
  {https://doi.org/10.1103/PhysRevB.81.134509} {\bibfield  {journal} {\bibinfo
  {journal} {Phys. Rev. B}\ }\textbf {\bibinfo {volume} {81}},\ \bibinfo
  {pages} {134509} (\bibinfo {year} {2010})}\BibitemShut {NoStop}%
\bibitem [{\citenamefont {Niu}\ \emph {et~al.}(2012)\citenamefont {Niu},
  \citenamefont {Chung}, \citenamefont {Hsu}, \citenamefont {Mandal},
  \citenamefont {Raghu},\ and\ \citenamefont {Chakravarty}}]{ipsita-sudip}%
  \BibitemOpen
  \bibfield  {author} {\bibinfo {author} {\bibfnamefont {Y.}~\bibnamefont
  {Niu}}, \bibinfo {author} {\bibfnamefont {S.~B.}\ \bibnamefont {Chung}},
  \bibinfo {author} {\bibfnamefont {C.-H.}\ \bibnamefont {Hsu}}, \bibinfo
  {author} {\bibfnamefont {I.}~\bibnamefont {Mandal}}, \bibinfo {author}
  {\bibfnamefont {S.}~\bibnamefont {Raghu}},\ and\ \bibinfo {author}
  {\bibfnamefont {S.}~\bibnamefont {Chakravarty}},\ }\bibfield  {title}
  {\bibinfo {title} {Majorana zero modes in a quantum ising chain with
  longer-ranged interactions},\ }\href
  {https://doi.org/10.1103/PhysRevB.85.035110} {\bibfield  {journal} {\bibinfo
  {journal} {Phys. Rev. B}\ }\textbf {\bibinfo {volume} {85}},\ \bibinfo
  {pages} {035110} (\bibinfo {year} {2012})}\BibitemShut {NoStop}%
\bibitem [{\citenamefont {Mandal}(2015)}]{Mandal2015}%
  \BibitemOpen
  \bibfield  {author} {\bibinfo {author} {\bibfnamefont {I.}~\bibnamefont
  {Mandal}},\ }\bibfield  {title} {\bibinfo {title} {Exceptional points for
  chiral majorana fermions in arbitrary dimensions},\ }\href
  {https://doi.org/http://dx.doi.org/10.1209/0295-5075/110/67005} {\bibfield
  {journal} {\bibinfo  {journal} {Europhysics Letters}\ }\textbf {\bibinfo
  {volume} {110}},\ \bibinfo {pages} {67005} (\bibinfo {year}
  {2015})}\BibitemShut {NoStop}%
\bibitem [{\citenamefont {Mandal}\ and\ \citenamefont
  {Tewari}(2016)}]{Mandal2016a}%
  \BibitemOpen
  \bibfield  {author} {\bibinfo {author} {\bibfnamefont {I.}~\bibnamefont
  {Mandal}}\ and\ \bibinfo {author} {\bibfnamefont {S.}~\bibnamefont
  {Tewari}},\ }\bibfield  {title} {\bibinfo {title} {Exceptional point
  description of one-dimensional chiral topological superconductors/superfluids
  in \{BDI\} class},\ }\href
  {https://doi.org/http://dx.doi.org/10.1016/j.physe.2015.12.009} {\bibfield
  {journal} {\bibinfo  {journal} {Physica E}\ }\textbf {\bibinfo {volume}
  {79}},\ \bibinfo {pages} {180 } (\bibinfo {year} {2016})}\BibitemShut
  {NoStop}%
\bibitem [{\citenamefont {Mandal}(2016)}]{Mandal2016b}%
  \BibitemOpen
  \bibfield  {author} {\bibinfo {author} {\bibfnamefont {I.}~\bibnamefont
  {Mandal}},\ }\bibfield  {title} {\bibinfo {title} {Counting majorana bound
  states using complex momenta},\ }\href
  {https://doi.org/http://dx.doi.org/10.5488/CMP.19.33703} {\bibfield
  {journal} {\bibinfo  {journal} {Condensed Matter Physics}\ }\textbf {\bibinfo
  {volume} {19}},\ \bibinfo {pages} {33703} (\bibinfo {year}
  {2016})}\BibitemShut {NoStop}%
\bibitem [{\citenamefont {Tewari}\ \emph {et~al.}(2011)\citenamefont {Tewari},
  \citenamefont {Stanescu}, \citenamefont {Sau},\ and\ \citenamefont
  {Sarma}}]{Tewari_2011}%
  \BibitemOpen
  \bibfield  {author} {\bibinfo {author} {\bibfnamefont {S.}~\bibnamefont
  {Tewari}}, \bibinfo {author} {\bibfnamefont {T.~D.}\ \bibnamefont
  {Stanescu}}, \bibinfo {author} {\bibfnamefont {J.~D.}\ \bibnamefont {Sau}},\
  and\ \bibinfo {author} {\bibfnamefont {S.~D.}\ \bibnamefont {Sarma}},\
  }\bibfield  {title} {\bibinfo {title} {Topologically non-trivial
  superconductivity in spin{\textendash}orbit-coupled systems: bulk phases and
  quantum phase transitions},\ }\href
  {https://doi.org/10.1088/1367-2630/13/6/065004} {\bibfield  {journal}
  {\bibinfo  {journal} {New Journal of Physics}\ }\textbf {\bibinfo {volume}
  {13}},\ \bibinfo {pages} {065004} (\bibinfo {year} {2011})}\BibitemShut
  {NoStop}%
\bibitem [{\citenamefont {Klinovaja}\ and\ \citenamefont
  {Loss}(2014)}]{jelena}%
  \BibitemOpen
  \bibfield  {author} {\bibinfo {author} {\bibfnamefont {J.}~\bibnamefont
  {Klinovaja}}\ and\ \bibinfo {author} {\bibfnamefont {D.}~\bibnamefont
  {Loss}},\ }\bibfield  {title} {\bibinfo {title} {Time-reversal invariant
  parafermions in interacting rashba nanowires},\ }\href
  {https://doi.org/10.1103/PhysRevB.90.045118} {\bibfield  {journal} {\bibinfo
  {journal} {Phys. Rev. B}\ }\textbf {\bibinfo {volume} {90}},\ \bibinfo
  {pages} {045118} (\bibinfo {year} {2014})}\BibitemShut {NoStop}%
\bibitem [{\citenamefont {Thakurathi}\ \emph {et~al.}(2018)\citenamefont
  {Thakurathi}, \citenamefont {Simon}, \citenamefont {Mandal}, \citenamefont
  {Klinovaja},\ and\ \citenamefont {Loss}}]{ips-kramers}%
  \BibitemOpen
  \bibfield  {author} {\bibinfo {author} {\bibfnamefont {M.}~\bibnamefont
  {Thakurathi}}, \bibinfo {author} {\bibfnamefont {P.}~\bibnamefont {Simon}},
  \bibinfo {author} {\bibfnamefont {I.}~\bibnamefont {Mandal}}, \bibinfo
  {author} {\bibfnamefont {J.}~\bibnamefont {Klinovaja}},\ and\ \bibinfo
  {author} {\bibfnamefont {D.}~\bibnamefont {Loss}},\ }\bibfield  {title}
  {\bibinfo {title} {Majorana kramers pairs in rashba double nanowires with
  interactions and disorder},\ }\href
  {https://doi.org/10.1103/PhysRevB.97.045415} {\bibfield  {journal} {\bibinfo
  {journal} {Phys. Rev. B}\ }\textbf {\bibinfo {volume} {97}},\ \bibinfo
  {pages} {045415} (\bibinfo {year} {2018})}\BibitemShut {NoStop}%
\bibitem [{\citenamefont {Sachdev}\ and\ \citenamefont {Ye}(1993)}]{sy}%
  \BibitemOpen
  \bibfield  {author} {\bibinfo {author} {\bibfnamefont {S.}~\bibnamefont
  {Sachdev}}\ and\ \bibinfo {author} {\bibfnamefont {J.}~\bibnamefont {Ye}},\
  }\bibfield  {title} {\bibinfo {title} {Gapless spin-fluid ground state in a
  random quantum heisenberg magnet},\ }\href
  {https://doi.org/10.1103/PhysRevLett.70.3339} {\bibfield  {journal} {\bibinfo
   {journal} {Phys. Rev. Lett.}\ }\textbf {\bibinfo {volume} {70}},\ \bibinfo
  {pages} {3339} (\bibinfo {year} {1993})}\BibitemShut {NoStop}%
\bibitem [{\citenamefont {{Kitaev}}(2015)}]{kitaev}%
  \BibitemOpen
  \bibfield  {author} {\bibinfo {author} {\bibfnamefont {A.}~\bibnamefont
  {{Kitaev}}},\ }\bibfield  {title} {\bibinfo {title} {A simple model of
  quantum holography},\ }\href
  {http://online.kitp.ucsb.edu/online/entangled15/} {\bibfield  {journal}
  {\bibinfo  {journal} {KITP strings seminar and Entanglement program}\ }
  (\bibinfo {year} {2015})}\BibitemShut {NoStop}%
\bibitem [{\citenamefont {Lantagne-Hurtubise}\ \emph
  {et~al.}(2018)\citenamefont {Lantagne-Hurtubise}, \citenamefont {Li},\ and\
  \citenamefont {Franz}}]{marcel}%
  \BibitemOpen
  \bibfield  {author} {\bibinfo {author} {\bibfnamefont {E.}~\bibnamefont
  {Lantagne-Hurtubise}}, \bibinfo {author} {\bibfnamefont {C.}~\bibnamefont
  {Li}},\ and\ \bibinfo {author} {\bibfnamefont {M.}~\bibnamefont {Franz}},\
  }\bibfield  {title} {\bibinfo {title} {Family of sachdev-ye-kitaev models
  motivated by experimental considerations},\ }\href
  {https://doi.org/10.1103/PhysRevB.97.235124} {\bibfield  {journal} {\bibinfo
  {journal} {Phys. Rev. B}\ }\textbf {\bibinfo {volume} {97}},\ \bibinfo
  {pages} {235124} (\bibinfo {year} {2018})}\BibitemShut {NoStop}%
\bibitem [{\citenamefont {Diez}\ \emph {et~al.}(2012)\citenamefont {Diez},
  \citenamefont {Dahlhaus}, \citenamefont {Wimmer},\ and\ \citenamefont
  {Beenakker}}]{beenakker}%
  \BibitemOpen
  \bibfield  {author} {\bibinfo {author} {\bibfnamefont {M.}~\bibnamefont
  {Diez}}, \bibinfo {author} {\bibfnamefont {J.~P.}\ \bibnamefont {Dahlhaus}},
  \bibinfo {author} {\bibfnamefont {M.}~\bibnamefont {Wimmer}},\ and\ \bibinfo
  {author} {\bibfnamefont {C.~W.~J.}\ \bibnamefont {Beenakker}},\ }\bibfield
  {title} {\bibinfo {title} {Andreev reflection from a topological
  superconductor with chiral symmetry},\ }\href
  {https://doi.org/10.1103/PhysRevB.86.094501} {\bibfield  {journal} {\bibinfo
  {journal} {Phys. Rev. B}\ }\textbf {\bibinfo {volume} {86}},\ \bibinfo
  {pages} {094501} (\bibinfo {year} {2012})}\BibitemShut {NoStop}%
\bibitem [{\citenamefont {Kaladzhyan}\ \emph {et~al.}(2017)\citenamefont
  {Kaladzhyan}, \citenamefont {Despres}, \citenamefont {Mandal},\ and\
  \citenamefont {Bena}}]{Kaladzhyan_2017}%
  \BibitemOpen
  \bibfield  {author} {\bibinfo {author} {\bibfnamefont {V.}~\bibnamefont
  {Kaladzhyan}}, \bibinfo {author} {\bibfnamefont {J.}~\bibnamefont {Despres}},
  \bibinfo {author} {\bibfnamefont {I.}~\bibnamefont {Mandal}},\ and\ \bibinfo
  {author} {\bibfnamefont {C.}~\bibnamefont {Bena}},\ }\bibfield  {title}
  {\bibinfo {title} {Majorana fermions in finite-size strips with in-plane
  magnetic fields},\ }\href {http://dx.doi.org/10.1140/epjb/e2017-80103-y}
  {\bibfield  {journal} {\bibinfo  {journal} {The European Physical Journal B}\
  }\textbf {\bibinfo {volume} {90}} (\bibinfo {year} {2017})}\BibitemShut
  {NoStop}%
\bibitem [{\citenamefont {Mercure}\ \emph {et~al.}(2012)\citenamefont
  {Mercure}, \citenamefont {Bangura}, \citenamefont {Xu}, \citenamefont
  {Wakeham}, \citenamefont {Carrington}, \citenamefont {Walmsley},
  \citenamefont {Greenblatt},\ and\ \citenamefont {Hussey}}]{limo1}%
  \BibitemOpen
  \bibfield  {author} {\bibinfo {author} {\bibfnamefont {J.-F.}\ \bibnamefont
  {Mercure}}, \bibinfo {author} {\bibfnamefont {A.~F.}\ \bibnamefont
  {Bangura}}, \bibinfo {author} {\bibfnamefont {X.}~\bibnamefont {Xu}},
  \bibinfo {author} {\bibfnamefont {N.}~\bibnamefont {Wakeham}}, \bibinfo
  {author} {\bibfnamefont {A.}~\bibnamefont {Carrington}}, \bibinfo {author}
  {\bibfnamefont {P.}~\bibnamefont {Walmsley}}, \bibinfo {author}
  {\bibfnamefont {M.}~\bibnamefont {Greenblatt}},\ and\ \bibinfo {author}
  {\bibfnamefont {N.~E.}\ \bibnamefont {Hussey}},\ }\bibfield  {title}
  {\bibinfo {title} {Upper critical magnetic field far above the paramagnetic
  pair-breaking limit of superconducting one-dimensional
  ${\mathrm{li}}_{0.9}{\mathrm{mo}}_{6}{\mathbf{o}}_{17}$ single crystals},\
  }\href {https://doi.org/10.1103/PhysRevLett.108.187003} {\bibfield  {journal}
  {\bibinfo  {journal} {Phys. Rev. Lett.}\ }\textbf {\bibinfo {volume} {108}},\
  \bibinfo {pages} {187003} (\bibinfo {year} {2012})}\BibitemShut {NoStop}%
\bibitem [{\citenamefont {Lebed}\ and\ \citenamefont
  {Sepper}(2013)}]{Lebed_13}%
  \BibitemOpen
  \bibfield  {author} {\bibinfo {author} {\bibfnamefont {A.~G.}\ \bibnamefont
  {Lebed}}\ and\ \bibinfo {author} {\bibfnamefont {O.}~\bibnamefont {Sepper}},\
  }\bibfield  {title} {\bibinfo {title} {Possible triplet superconductivity in
  the quasi-one-dimensional conductor li${}_{0.9}$mo${}_{6}$o${}_{17}$},\
  }\href {https://doi.org/10.1103/PhysRevB.87.100511} {\bibfield  {journal}
  {\bibinfo  {journal} {Phys. Rev. B}\ }\textbf {\bibinfo {volume} {87}},\
  \bibinfo {pages} {100511} (\bibinfo {year} {2013})}\BibitemShut {NoStop}%
\bibitem [{\citenamefont {Lebed}\ \emph {et~al.}(2000)\citenamefont {Lebed},
  \citenamefont {Machida},\ and\ \citenamefont {Ozaki}}]{Lebed_00}%
  \BibitemOpen
  \bibfield  {author} {\bibinfo {author} {\bibfnamefont {A.~G.}\ \bibnamefont
  {Lebed}}, \bibinfo {author} {\bibfnamefont {K.}~\bibnamefont {Machida}},\
  and\ \bibinfo {author} {\bibfnamefont {M.}~\bibnamefont {Ozaki}},\ }\bibfield
   {title} {\bibinfo {title} {Triplet electron pairing and anisotropic spin
  susceptibility in organic superconductors
  $({\mathrm{t}\mathrm{m}\mathrm{t}\mathrm{s}\mathrm{f})}_{2}x$},\ }\href
  {https://doi.org/10.1103/PhysRevB.62.R795} {\bibfield  {journal} {\bibinfo
  {journal} {Phys. Rev. B}\ }\textbf {\bibinfo {volume} {62}},\ \bibinfo
  {pages} {R795} (\bibinfo {year} {2000})}\BibitemShut {NoStop}%
\bibitem [{\citenamefont {Lee}\ \emph {et~al.}(2003)\citenamefont {Lee},
  \citenamefont {Chow}, \citenamefont {Clark}, \citenamefont {Strouse},
  \citenamefont {Naughton}, \citenamefont {Chaikin},\ and\ \citenamefont
  {Brown}}]{triplet2}%
  \BibitemOpen
  \bibfield  {author} {\bibinfo {author} {\bibfnamefont {I.~J.}\ \bibnamefont
  {Lee}}, \bibinfo {author} {\bibfnamefont {D.~S.}\ \bibnamefont {Chow}},
  \bibinfo {author} {\bibfnamefont {W.~G.}\ \bibnamefont {Clark}}, \bibinfo
  {author} {\bibfnamefont {M.~J.}\ \bibnamefont {Strouse}}, \bibinfo {author}
  {\bibfnamefont {M.~J.}\ \bibnamefont {Naughton}}, \bibinfo {author}
  {\bibfnamefont {P.~M.}\ \bibnamefont {Chaikin}},\ and\ \bibinfo {author}
  {\bibfnamefont {S.~E.}\ \bibnamefont {Brown}},\ }\bibfield  {title} {\bibinfo
  {title} {Evidence from ${}^{77}\mathrm{Se}$ knight shifts for triplet
  superconductivity in $(\mathrm{TMTSF}{)}_{2}{\mathrm{pf}}_{6}$},\ }\href
  {https://doi.org/10.1103/PhysRevB.68.092510} {\bibfield  {journal} {\bibinfo
  {journal} {Phys. Rev. B}\ }\textbf {\bibinfo {volume} {68}},\ \bibinfo
  {pages} {092510} (\bibinfo {year} {2003})}\BibitemShut {NoStop}%
\bibitem [{\citenamefont {Shinagawa}\ \emph {et~al.}(2007)\citenamefont
  {Shinagawa}, \citenamefont {Kurosaki}, \citenamefont {Zhang}, \citenamefont
  {Parker}, \citenamefont {Brown}, \citenamefont {J\'erome}, \citenamefont
  {Christensen},\ and\ \citenamefont {Bechgaard}}]{Shinagawa_2007}%
  \BibitemOpen
  \bibfield  {author} {\bibinfo {author} {\bibfnamefont {J.}~\bibnamefont
  {Shinagawa}}, \bibinfo {author} {\bibfnamefont {Y.}~\bibnamefont {Kurosaki}},
  \bibinfo {author} {\bibfnamefont {F.}~\bibnamefont {Zhang}}, \bibinfo
  {author} {\bibfnamefont {C.}~\bibnamefont {Parker}}, \bibinfo {author}
  {\bibfnamefont {S.~E.}\ \bibnamefont {Brown}}, \bibinfo {author}
  {\bibfnamefont {D.}~\bibnamefont {J\'erome}}, \bibinfo {author}
  {\bibfnamefont {J.~B.}\ \bibnamefont {Christensen}},\ and\ \bibinfo {author}
  {\bibfnamefont {K.}~\bibnamefont {Bechgaard}},\ }\bibfield  {title} {\bibinfo
  {title} {Superconducting state of the organic conductor
  $(\mathrm{TMTSF}{)}_{2}{\mathrm{clo}}_{4}$},\ }\href
  {https://doi.org/10.1103/PhysRevLett.98.147002} {\bibfield  {journal}
  {\bibinfo  {journal} {Phys. Rev. Lett.}\ }\textbf {\bibinfo {volume} {98}},\
  \bibinfo {pages} {147002} (\bibinfo {year} {2007})}\BibitemShut {NoStop}%
\bibitem [{\citenamefont {Raghu}\ \emph {et~al.}(2010)\citenamefont {Raghu},
  \citenamefont {Kapitulnik},\ and\ \citenamefont {Kivelson}}]{raghu}%
  \BibitemOpen
  \bibfield  {author} {\bibinfo {author} {\bibfnamefont {S.}~\bibnamefont
  {Raghu}}, \bibinfo {author} {\bibfnamefont {A.}~\bibnamefont {Kapitulnik}},\
  and\ \bibinfo {author} {\bibfnamefont {S.~A.}\ \bibnamefont {Kivelson}},\
  }\bibfield  {title} {\bibinfo {title} {Hidden quasi-one-dimensional
  superconductivity in ${\mathrm{sr}}_{2}{\mathrm{ruo}}_{4}$},\ }\href
  {https://doi.org/10.1103/PhysRevLett.105.136401} {\bibfield  {journal}
  {\bibinfo  {journal} {Phys. Rev. Lett.}\ }\textbf {\bibinfo {volume} {105}},\
  \bibinfo {pages} {136401} (\bibinfo {year} {2010})}\BibitemShut {NoStop}%
\bibitem [{\citenamefont {Sengupta}\ \emph {et~al.}(2001)\citenamefont
  {Sengupta}, \citenamefont {\ifmmode \check{Z}\else
  \v{Z}\fi{}uti\ifmmode~\acute{c}\else \'{c}\fi{}}, \citenamefont {Kwon},
  \citenamefont {Yakovenko},\ and\ \citenamefont {Das~Sarma}}]{Yakovenko}%
  \BibitemOpen
  \bibfield  {author} {\bibinfo {author} {\bibfnamefont {K.}~\bibnamefont
  {Sengupta}}, \bibinfo {author} {\bibfnamefont {I.}~\bibnamefont {\ifmmode
  \check{Z}\else \v{Z}\fi{}uti\ifmmode~\acute{c}\else \'{c}\fi{}}}, \bibinfo
  {author} {\bibfnamefont {H.-J.}\ \bibnamefont {Kwon}}, \bibinfo {author}
  {\bibfnamefont {V.~M.}\ \bibnamefont {Yakovenko}},\ and\ \bibinfo {author}
  {\bibfnamefont {S.}~\bibnamefont {Das~Sarma}},\ }\bibfield  {title} {\bibinfo
  {title} {Midgap edge states and pairing symmetry of quasi-one-dimensional
  organic superconductors},\ }\href
  {https://doi.org/10.1103/PhysRevB.63.144531} {\bibfield  {journal} {\bibinfo
  {journal} {Phys. Rev. B}\ }\textbf {\bibinfo {volume} {63}},\ \bibinfo
  {pages} {144531} (\bibinfo {year} {2001})}\BibitemShut {NoStop}%
\bibitem [{\citenamefont {Dumitrescu}\ and\ \citenamefont
  {Tewari}(2013)}]{Dumitrescu13}%
  \BibitemOpen
  \bibfield  {author} {\bibinfo {author} {\bibfnamefont {E.}~\bibnamefont
  {Dumitrescu}}\ and\ \bibinfo {author} {\bibfnamefont {S.}~\bibnamefont
  {Tewari}},\ }\bibfield  {title} {\bibinfo {title} {Topological properties of
  the time-reversal-symmetric kitaev chain and applications to organic
  superconductors},\ }\href {https://doi.org/10.1103/PhysRevB.88.220505}
  {\bibfield  {journal} {\bibinfo  {journal} {Phys. Rev. B}\ }\textbf {\bibinfo
  {volume} {88}},\ \bibinfo {pages} {220505} (\bibinfo {year}
  {2013})}\BibitemShut {NoStop}%
\bibitem [{\citenamefont {Ikegaya}\ \emph {et~al.}(2016)\citenamefont
  {Ikegaya}, \citenamefont {Suzuki}, \citenamefont {Tanaka},\ and\
  \citenamefont {Asano}}]{yukio1}%
  \BibitemOpen
  \bibfield  {author} {\bibinfo {author} {\bibfnamefont {S.}~\bibnamefont
  {Ikegaya}}, \bibinfo {author} {\bibfnamefont {S.-I.}\ \bibnamefont {Suzuki}},
  \bibinfo {author} {\bibfnamefont {Y.}~\bibnamefont {Tanaka}},\ and\ \bibinfo
  {author} {\bibfnamefont {Y.}~\bibnamefont {Asano}},\ }\bibfield  {title}
  {\bibinfo {title} {Quantization of conductance minimum and index theorem},\
  }\href {https://doi.org/10.1103/PhysRevB.94.054512} {\bibfield  {journal}
  {\bibinfo  {journal} {Phys. Rev. B}\ }\textbf {\bibinfo {volume} {94}},\
  \bibinfo {pages} {054512} (\bibinfo {year} {2016})}\BibitemShut {NoStop}%
\bibitem [{\citenamefont {Tanaka}\ and\ \citenamefont
  {Golubov}(2007)}]{yukio2}%
  \BibitemOpen
  \bibfield  {author} {\bibinfo {author} {\bibfnamefont {Y.}~\bibnamefont
  {Tanaka}}\ and\ \bibinfo {author} {\bibfnamefont {A.~A.}\ \bibnamefont
  {Golubov}},\ }\bibfield  {title} {\bibinfo {title} {Theory of the proximity
  effect in junctions with unconventional superconductors},\ }\href
  {https://doi.org/10.1103/PhysRevLett.98.037003} {\bibfield  {journal}
  {\bibinfo  {journal} {Phys. Rev. Lett.}\ }\textbf {\bibinfo {volume} {98}},\
  \bibinfo {pages} {037003} (\bibinfo {year} {2007})}\BibitemShut {NoStop}%
\bibitem [{\citenamefont {Cayao}\ \emph {et~al.}(2017)\citenamefont {Cayao},
  \citenamefont {San-Jose}, \citenamefont {Black-Schaffer}, \citenamefont
  {Aguado},\ and\ \citenamefont {Prada}}]{jorge1}%
  \BibitemOpen
  \bibfield  {author} {\bibinfo {author} {\bibfnamefont {J.}~\bibnamefont
  {Cayao}}, \bibinfo {author} {\bibfnamefont {P.}~\bibnamefont {San-Jose}},
  \bibinfo {author} {\bibfnamefont {A.~M.}\ \bibnamefont {Black-Schaffer}},
  \bibinfo {author} {\bibfnamefont {R.}~\bibnamefont {Aguado}},\ and\ \bibinfo
  {author} {\bibfnamefont {E.}~\bibnamefont {Prada}},\ }\bibfield  {title}
  {\bibinfo {title} {Majorana splitting from critical currents in josephson
  junctions},\ }\href {https://doi.org/10.1103/PhysRevB.96.205425} {\bibfield
  {journal} {\bibinfo  {journal} {Phys. Rev. B}\ }\textbf {\bibinfo {volume}
  {96}},\ \bibinfo {pages} {205425} (\bibinfo {year} {2017})}\BibitemShut
  {NoStop}%
\bibitem [{\citenamefont {{Cayao}}\ \emph {et~al.}(2018)\citenamefont
  {{Cayao}}, \citenamefont {{Black-Schaffer}}, \citenamefont {{Prada}},\ and\
  \citenamefont {{Aguado}}}]{jorge2}%
  \BibitemOpen
  \bibfield  {author} {\bibinfo {author} {\bibfnamefont {J.}~\bibnamefont
  {{Cayao}}}, \bibinfo {author} {\bibfnamefont {A.~M.}\ \bibnamefont
  {{Black-Schaffer}}}, \bibinfo {author} {\bibfnamefont {E.}~\bibnamefont
  {{Prada}}},\ and\ \bibinfo {author} {\bibfnamefont {R.}~\bibnamefont
  {{Aguado}}},\ }\bibfield  {title} {\bibinfo {title} {{Andreev spectrum and
  supercurrents in nanowire-based SNS junctions containing Majorana bound
  states}},\ }\href {https://doi.org/10.3762/bjnano.9.127} {\bibfield
  {journal} {\bibinfo  {journal} {Beilstein J. Nanotechnol.}\ }\textbf
  {\bibinfo {volume} {9}},\ \bibinfo {pages} {1339} (\bibinfo {year}
  {2018})}\BibitemShut {NoStop}%
\bibitem [{\citenamefont {Janvier}\ \emph {et~al.}(2015)\citenamefont
  {Janvier}, \citenamefont {Tosi}, \citenamefont {Bretheau}, \citenamefont
  {Girit}, \citenamefont {Stern}, \citenamefont {Bertet}, \citenamefont
  {Joyez}, \citenamefont {Vion}, \citenamefont {Esteve}, \citenamefont
  {Goffman}, \citenamefont {Pothier},\ and\ \citenamefont {Urbina}}]{urbina}%
  \BibitemOpen
  \bibfield  {author} {\bibinfo {author} {\bibfnamefont {C.}~\bibnamefont
  {Janvier}}, \bibinfo {author} {\bibfnamefont {L.}~\bibnamefont {Tosi}},
  \bibinfo {author} {\bibfnamefont {L.}~\bibnamefont {Bretheau}}, \bibinfo
  {author} {\bibfnamefont {{\c C}.~{\"O}.}\ \bibnamefont {Girit}}, \bibinfo
  {author} {\bibfnamefont {M.}~\bibnamefont {Stern}}, \bibinfo {author}
  {\bibfnamefont {P.}~\bibnamefont {Bertet}}, \bibinfo {author} {\bibfnamefont
  {P.}~\bibnamefont {Joyez}}, \bibinfo {author} {\bibfnamefont
  {D.}~\bibnamefont {Vion}}, \bibinfo {author} {\bibfnamefont {D.}~\bibnamefont
  {Esteve}}, \bibinfo {author} {\bibfnamefont {M.~F.}\ \bibnamefont {Goffman}},
  \bibinfo {author} {\bibfnamefont {H.}~\bibnamefont {Pothier}},\ and\ \bibinfo
  {author} {\bibfnamefont {C.}~\bibnamefont {Urbina}},\ }\bibfield  {title}
  {\bibinfo {title} {Coherent manipulation of andreev states in superconducting
  atomic contacts},\ }\href {https://doi.org/10.1126/science.aab2179} {\
  \textbf {\bibinfo {volume} {349}},\ \bibinfo {pages} {1199} (\bibinfo {year}
  {2015})}\BibitemShut {NoStop}%
\bibitem [{\citenamefont {Wallraff}\ \emph {et~al.}(2007)\citenamefont
  {Wallraff}, \citenamefont {Schuster}, \citenamefont {Blais}, \citenamefont
  {Gambetta}, \citenamefont {Schreier}, \citenamefont {Frunzio}, \citenamefont
  {Devoret}, \citenamefont {Girvin},\ and\ \citenamefont
  {Schoelkopf}}]{Devoret}%
  \BibitemOpen
  \bibfield  {author} {\bibinfo {author} {\bibfnamefont {A.}~\bibnamefont
  {Wallraff}}, \bibinfo {author} {\bibfnamefont {D.~I.}\ \bibnamefont
  {Schuster}}, \bibinfo {author} {\bibfnamefont {A.}~\bibnamefont {Blais}},
  \bibinfo {author} {\bibfnamefont {J.~M.}\ \bibnamefont {Gambetta}}, \bibinfo
  {author} {\bibfnamefont {J.}~\bibnamefont {Schreier}}, \bibinfo {author}
  {\bibfnamefont {L.}~\bibnamefont {Frunzio}}, \bibinfo {author} {\bibfnamefont
  {M.~H.}\ \bibnamefont {Devoret}}, \bibinfo {author} {\bibfnamefont {S.~M.}\
  \bibnamefont {Girvin}},\ and\ \bibinfo {author} {\bibfnamefont {R.~J.}\
  \bibnamefont {Schoelkopf}},\ }\bibfield  {title} {\bibinfo {title} {Sideband
  transitions and two-tone spectroscopy of a superconducting qubit strongly
  coupled to an on-chip cavity},\ }\href
  {https://doi.org/10.1103/PhysRevLett.99.050501} {\bibfield  {journal}
  {\bibinfo  {journal} {Phys. Rev. Lett.}\ }\textbf {\bibinfo {volume} {99}},\
  \bibinfo {pages} {050501} (\bibinfo {year} {2007})}\BibitemShut {NoStop}%
\bibitem [{\citenamefont {Groth}\ \emph {et~al.}(2014)\citenamefont {Groth},
  \citenamefont {Wimmer}, \citenamefont {Akhmerov},\ and\ \citenamefont
  {Waintal}}]{Groth_2014}%
  \BibitemOpen
  \bibfield  {author} {\bibinfo {author} {\bibfnamefont {C.~W.}\ \bibnamefont
  {Groth}}, \bibinfo {author} {\bibfnamefont {M.}~\bibnamefont {Wimmer}},
  \bibinfo {author} {\bibfnamefont {A.~R.}\ \bibnamefont {Akhmerov}},\ and\
  \bibinfo {author} {\bibfnamefont {X.}~\bibnamefont {Waintal}},\ }\bibfield
  {title} {\bibinfo {title} {Kwant: a software package for quantum transport},\
  }\href {https://doi.org/10.1088/1367-2630/16/6/063065} {\bibfield  {journal}
  {\bibinfo  {journal} {New Journal of Physics}\ }\textbf {\bibinfo {volume}
  {16}},\ \bibinfo {pages} {063065} (\bibinfo {year} {2014})}\BibitemShut
  {NoStop}%
\bibitem [{\citenamefont {Tewari}\ and\ \citenamefont {Sau}(2012)}]{tewsau}%
  \BibitemOpen
  \bibfield  {author} {\bibinfo {author} {\bibfnamefont {S.}~\bibnamefont
  {Tewari}}\ and\ \bibinfo {author} {\bibfnamefont {J.~D.}\ \bibnamefont
  {Sau}},\ }\bibfield  {title} {\bibinfo {title} {Topological invariants for
  spin-orbit coupled superconductor nanowires},\ }\href
  {https://doi.org/10.1103/PhysRevLett.109.150408} {\bibfield  {journal}
  {\bibinfo  {journal} {Phys. Rev. Lett.}\ }\textbf {\bibinfo {volume} {109}},\
  \bibinfo {pages} {150408} (\bibinfo {year} {2012})}\BibitemShut {NoStop}%
\bibitem [{\citenamefont {{de Gennes}}\ and\ \citenamefont
  {Saint-James}(1963)}]{DEGENNES}%
  \BibitemOpen
  \bibfield  {author} {\bibinfo {author} {\bibfnamefont {P.}~\bibnamefont {{de
  Gennes}}}\ and\ \bibinfo {author} {\bibfnamefont {D.}~\bibnamefont
  {Saint-James}},\ }\bibfield  {title} {\bibinfo {title} {Elementary
  excitations in the vicinity of a normal metal-superconducting metal
  contact},\ }\href
  {https://doi.org/https://doi.org/10.1016/0031-9163(63)90148-3} {\bibfield
  {journal} {\bibinfo  {journal} {Physics Letters}\ }\textbf {\bibinfo {volume}
  {4}},\ \bibinfo {pages} {151 } (\bibinfo {year} {1963})}\BibitemShut
  {NoStop}%
\bibitem [{\citenamefont {Kells}\ \emph {et~al.}(2012)\citenamefont {Kells},
  \citenamefont {Meidan},\ and\ \citenamefont {Brouwer}}]{PhysRevB.86.100503}%
  \BibitemOpen
  \bibfield  {author} {\bibinfo {author} {\bibfnamefont {G.}~\bibnamefont
  {Kells}}, \bibinfo {author} {\bibfnamefont {D.}~\bibnamefont {Meidan}},\ and\
  \bibinfo {author} {\bibfnamefont {P.~W.}\ \bibnamefont {Brouwer}},\
  }\bibfield  {title} {\bibinfo {title} {Near-zero-energy end states in
  topologically trivial spin-orbit coupled superconducting nanowires with a
  smooth confinement},\ }\href {https://doi.org/10.1103/PhysRevB.86.100503}
  {\bibfield  {journal} {\bibinfo  {journal} {Phys. Rev. B}\ }\textbf {\bibinfo
  {volume} {86}},\ \bibinfo {pages} {100503} (\bibinfo {year}
  {2012})}\BibitemShut {NoStop}%
\bibitem [{\citenamefont {Prada}\ \emph {et~al.}(2012)\citenamefont {Prada},
  \citenamefont {San-Jose},\ and\ \citenamefont {Aguado}}]{PhysRevB.86.180503}%
  \BibitemOpen
  \bibfield  {author} {\bibinfo {author} {\bibfnamefont {E.}~\bibnamefont
  {Prada}}, \bibinfo {author} {\bibfnamefont {P.}~\bibnamefont {San-Jose}},\
  and\ \bibinfo {author} {\bibfnamefont {R.}~\bibnamefont {Aguado}},\
  }\bibfield  {title} {\bibinfo {title} {Transport spectroscopy of $ns$
  nanowire junctions with majorana fermions},\ }\href
  {https://doi.org/10.1103/PhysRevB.86.180503} {\bibfield  {journal} {\bibinfo
  {journal} {Phys. Rev. B}\ }\textbf {\bibinfo {volume} {86}},\ \bibinfo
  {pages} {180503} (\bibinfo {year} {2012})}\BibitemShut {NoStop}%
\bibitem [{\citenamefont {Moore}\ \emph
  {et~al.}(2018{\natexlab{a}})\citenamefont {Moore}, \citenamefont {Stanescu},\
  and\ \citenamefont {Tewari}}]{PhysRevB.97.165302}%
  \BibitemOpen
  \bibfield  {author} {\bibinfo {author} {\bibfnamefont {C.}~\bibnamefont
  {Moore}}, \bibinfo {author} {\bibfnamefont {T.~D.}\ \bibnamefont
  {Stanescu}},\ and\ \bibinfo {author} {\bibfnamefont {S.}~\bibnamefont
  {Tewari}},\ }\bibfield  {title} {\bibinfo {title} {Two-terminal charge
  tunneling: Disentangling majorana zero modes from partially separated andreev
  bound states in semiconductor-superconductor heterostructures},\ }\href
  {https://doi.org/10.1103/PhysRevB.97.165302} {\bibfield  {journal} {\bibinfo
  {journal} {Phys. Rev. B}\ }\textbf {\bibinfo {volume} {97}},\ \bibinfo
  {pages} {165302} (\bibinfo {year} {2018}{\natexlab{a}})}\BibitemShut
  {NoStop}%
\bibitem [{\citenamefont {Liu}\ \emph {et~al.}(2017)\citenamefont {Liu},
  \citenamefont {Sau}, \citenamefont {Stanescu},\ and\ \citenamefont
  {Das~Sarma}}]{PhysRevB.96.075161}%
  \BibitemOpen
  \bibfield  {author} {\bibinfo {author} {\bibfnamefont {C.-X.}\ \bibnamefont
  {Liu}}, \bibinfo {author} {\bibfnamefont {J.~D.}\ \bibnamefont {Sau}},
  \bibinfo {author} {\bibfnamefont {T.~D.}\ \bibnamefont {Stanescu}},\ and\
  \bibinfo {author} {\bibfnamefont {S.}~\bibnamefont {Das~Sarma}},\ }\bibfield
  {title} {\bibinfo {title} {Andreev bound states versus majorana bound states
  in quantum dot-nanowire-superconductor hybrid structures: Trivial versus
  topological zero-bias conductance peaks},\ }\href
  {https://doi.org/10.1103/PhysRevB.96.075161} {\bibfield  {journal} {\bibinfo
  {journal} {Phys. Rev. B}\ }\textbf {\bibinfo {volume} {96}},\ \bibinfo
  {pages} {075161} (\bibinfo {year} {2017})}\BibitemShut {NoStop}%
\bibitem [{\citenamefont {Moore}\ \emph
  {et~al.}(2018{\natexlab{b}})\citenamefont {Moore}, \citenamefont {Zeng},
  \citenamefont {Stanescu},\ and\ \citenamefont {Tewari}}]{PhysRevB.98.155314}%
  \BibitemOpen
  \bibfield  {author} {\bibinfo {author} {\bibfnamefont {C.}~\bibnamefont
  {Moore}}, \bibinfo {author} {\bibfnamefont {C.}~\bibnamefont {Zeng}},
  \bibinfo {author} {\bibfnamefont {T.~D.}\ \bibnamefont {Stanescu}},\ and\
  \bibinfo {author} {\bibfnamefont {S.}~\bibnamefont {Tewari}},\ }\bibfield
  {title} {\bibinfo {title} {Quantized zero-bias conductance plateau in
  semiconductor-superconductor heterostructures without topological majorana
  zero modes},\ }\href {https://doi.org/10.1103/PhysRevB.98.155314} {\bibfield
  {journal} {\bibinfo  {journal} {Phys. Rev. B}\ }\textbf {\bibinfo {volume}
  {98}},\ \bibinfo {pages} {155314} (\bibinfo {year}
  {2018}{\natexlab{b}})}\BibitemShut {NoStop}%
\bibitem [{\citenamefont {Vuik}\ \emph {et~al.}(2019)\citenamefont {Vuik},
  \citenamefont {Nijholt}, \citenamefont {Akhmerov},\ and\ \citenamefont
  {Wimmer}}]{Vuik_2019}%
  \BibitemOpen
  \bibfield  {author} {\bibinfo {author} {\bibfnamefont {A.}~\bibnamefont
  {Vuik}}, \bibinfo {author} {\bibfnamefont {B.}~\bibnamefont {Nijholt}},
  \bibinfo {author} {\bibfnamefont {A.}~\bibnamefont {Akhmerov}},\ and\
  \bibinfo {author} {\bibfnamefont {M.}~\bibnamefont {Wimmer}},\ }\bibfield
  {title} {\bibinfo {title} {Reproducing topological properties with
  quasi-majorana states},\ }\bibfield  {journal} {\bibinfo  {journal} {SciPost
  Physics}\ }\textbf {\bibinfo {volume} {7}},\ \href
  {https://doi.org/10.21468/scipostphys.7.5.061} {10.21468/scipostphys.7.5.061}
  (\bibinfo {year} {2019})\BibitemShut {NoStop}%
\bibitem [{\citenamefont {Stanescu}\ \emph {et~al.}(2011)\citenamefont
  {Stanescu}, \citenamefont {Lutchyn},\ and\ \citenamefont
  {Das~Sarma}}]{PhysRevB.84.144522}%
  \BibitemOpen
  \bibfield  {author} {\bibinfo {author} {\bibfnamefont {T.~D.}\ \bibnamefont
  {Stanescu}}, \bibinfo {author} {\bibfnamefont {R.~M.}\ \bibnamefont
  {Lutchyn}},\ and\ \bibinfo {author} {\bibfnamefont {S.}~\bibnamefont
  {Das~Sarma}},\ }\bibfield  {title} {\bibinfo {title} {Majorana fermions in
  semiconductor nanowires},\ }\href
  {https://doi.org/10.1103/PhysRevB.84.144522} {\bibfield  {journal} {\bibinfo
  {journal} {Phys. Rev. B}\ }\textbf {\bibinfo {volume} {84}},\ \bibinfo
  {pages} {144522} (\bibinfo {year} {2011})}\BibitemShut {NoStop}%
\bibitem [{\citenamefont {Lee}\ \emph {et~al.}(2001)\citenamefont {Lee},
  \citenamefont {Brown}, \citenamefont {Clark}, \citenamefont {Strouse},
  \citenamefont {Naughton}, \citenamefont {Kang},\ and\ \citenamefont
  {Chaikin}}]{triplet1}%
  \BibitemOpen
  \bibfield  {author} {\bibinfo {author} {\bibfnamefont {I.~J.}\ \bibnamefont
  {Lee}}, \bibinfo {author} {\bibfnamefont {S.~E.}\ \bibnamefont {Brown}},
  \bibinfo {author} {\bibfnamefont {W.~G.}\ \bibnamefont {Clark}}, \bibinfo
  {author} {\bibfnamefont {M.~J.}\ \bibnamefont {Strouse}}, \bibinfo {author}
  {\bibfnamefont {M.~J.}\ \bibnamefont {Naughton}}, \bibinfo {author}
  {\bibfnamefont {W.}~\bibnamefont {Kang}},\ and\ \bibinfo {author}
  {\bibfnamefont {P.~M.}\ \bibnamefont {Chaikin}},\ }\bibfield  {title}
  {\bibinfo {title} {Triplet superconductivity in an organic superconductor
  probed by nmr knight shift},\ }\href
  {https://doi.org/10.1103/PhysRevLett.88.017004} {\bibfield  {journal}
  {\bibinfo  {journal} {Phys. Rev. Lett.}\ }\textbf {\bibinfo {volume} {88}},\
  \bibinfo {pages} {017004} (\bibinfo {year} {2001})}\BibitemShut {NoStop}%
\bibitem [{\citenamefont {Dumitrescu}\ \emph {et~al.}(2015)\citenamefont
  {Dumitrescu}, \citenamefont {Stanescu},\ and\ \citenamefont
  {Tewari}}]{tudor15}%
  \BibitemOpen
  \bibfield  {author} {\bibinfo {author} {\bibfnamefont {E.}~\bibnamefont
  {Dumitrescu}}, \bibinfo {author} {\bibfnamefont {T.~D.}\ \bibnamefont
  {Stanescu}},\ and\ \bibinfo {author} {\bibfnamefont {S.}~\bibnamefont
  {Tewari}},\ }\bibfield  {title} {\bibinfo {title} {Hidden-symmetry decoupling
  of majorana bound states in topological superconductors},\ }\href
  {https://doi.org/10.1103/PhysRevB.91.121413} {\bibfield  {journal} {\bibinfo
  {journal} {Phys. Rev. B}\ }\textbf {\bibinfo {volume} {91}},\ \bibinfo
  {pages} {121413} (\bibinfo {year} {2015})}\BibitemShut {NoStop}%
\bibitem [{\citenamefont {Kitaev}(2001)}]{Kitaev_2001}%
  \BibitemOpen
  \bibfield  {author} {\bibinfo {author} {\bibfnamefont {A.~Y.}\ \bibnamefont
  {Kitaev}},\ }\bibfield  {title} {\bibinfo {title} {Unpaired majorana fermions
  in quantum wires},\ }\href {https://doi.org/10.1070/1063-7869/44/10s/s29}
  {\bibfield  {journal} {\bibinfo  {journal} {Physics-Uspekhi}\ }\textbf
  {\bibinfo {volume} {44}},\ \bibinfo {pages} {131} (\bibinfo {year}
  {2001})}\BibitemShut {NoStop}%
\bibitem [{\citenamefont {Nijholt}()}]{pf}%
  \BibitemOpen
  \bibfield  {author} {\bibinfo {author} {\bibfnamefont {B.}~\bibnamefont
  {Nijholt}},\ }\href
  {https://gitlab.kwant-project.org/qt/topocm/blob/9b8af94790ddf8966f031065a248ab4ec68ed19d/code/pfaffian.py}
  {\emph {\bibinfo {title} {{Pfaffian.py}}}}\BibitemShut {NoStop}%
\bibitem [{\citenamefont {Budich}\ and\ \citenamefont {Ardonne}(2013)}]{DIII}%
  \BibitemOpen
  \bibfield  {author} {\bibinfo {author} {\bibfnamefont {J.~C.}\ \bibnamefont
  {Budich}}\ and\ \bibinfo {author} {\bibfnamefont {E.}~\bibnamefont
  {Ardonne}},\ }\bibfield  {title} {\bibinfo {title} {Topological invariant for
  generic one-dimensional time-reversal-symmetric superconductors in class
  diii},\ }\href {https://doi.org/10.1103/PhysRevB.88.134523} {\bibfield
  {journal} {\bibinfo  {journal} {Phys. Rev. B}\ }\textbf {\bibinfo {volume}
  {88}},\ \bibinfo {pages} {134523} (\bibinfo {year} {2013})}\BibitemShut
  {NoStop}%
\bibitem [{\citenamefont {{Pientka}}\ \emph {et~al.}(2013)\citenamefont
  {{Pientka}}, \citenamefont {{Romito}}, \citenamefont {{Duckheim}},
  \citenamefont {{Oreg}},\ and\ \citenamefont {{von Oppen}}}]{felix}%
  \BibitemOpen
  \bibfield  {author} {\bibinfo {author} {\bibfnamefont {F.}~\bibnamefont
  {{Pientka}}}, \bibinfo {author} {\bibfnamefont {A.}~\bibnamefont {{Romito}}},
  \bibinfo {author} {\bibfnamefont {M.}~\bibnamefont {{Duckheim}}}, \bibinfo
  {author} {\bibfnamefont {Y.}~\bibnamefont {{Oreg}}},\ and\ \bibinfo {author}
  {\bibfnamefont {F.}~\bibnamefont {{von Oppen}}},\ }\bibfield  {title}
  {\bibinfo {title} {{Signatures of topological phase transitions in mesoscopic
  superconducting rings}},\ }\href
  {https://doi.org/10.1088/1367-2630/15/2/025001} {\bibfield  {journal}
  {\bibinfo  {journal} {New Journal of Physics}\ }\textbf {\bibinfo {volume}
  {15}},\ \bibinfo {eid} {025001} (\bibinfo {year} {2013})}\BibitemShut
  {NoStop}%
\bibitem [{\citenamefont {van Woerkom}\ \emph {et~al.}(2017)\citenamefont {van
  Woerkom}, \citenamefont {Proutski}, \citenamefont {van Heck}, \citenamefont
  {Bouman}, \citenamefont {Väyrynen}, \citenamefont {Glazman}, \citenamefont
  {Krogstrup}, \citenamefont {Nygård}, \citenamefont {Kouwenhoven},\ and\
  \citenamefont {Geresdi}}]{van_Woerkom}%
  \BibitemOpen
  \bibfield  {author} {\bibinfo {author} {\bibfnamefont {D.~J.}\ \bibnamefont
  {van Woerkom}}, \bibinfo {author} {\bibfnamefont {A.}~\bibnamefont
  {Proutski}}, \bibinfo {author} {\bibfnamefont {B.}~\bibnamefont {van Heck}},
  \bibinfo {author} {\bibfnamefont {D.}~\bibnamefont {Bouman}}, \bibinfo
  {author} {\bibfnamefont {J.~I.}\ \bibnamefont {Väyrynen}}, \bibinfo {author}
  {\bibfnamefont {L.~I.}\ \bibnamefont {Glazman}}, \bibinfo {author}
  {\bibfnamefont {P.}~\bibnamefont {Krogstrup}}, \bibinfo {author}
  {\bibfnamefont {J.}~\bibnamefont {Nygård}}, \bibinfo {author} {\bibfnamefont
  {L.~P.}\ \bibnamefont {Kouwenhoven}},\ and\ \bibinfo {author} {\bibfnamefont
  {A.}~\bibnamefont {Geresdi}},\ }\bibfield  {title} {\bibinfo {title}
  {Microwave spectroscopy of spinful andreev bound states in ballistic
  semiconductor josephson junctions},\ }\href
  {https://doi.org/10.1038/nphys4150} {\bibfield  {journal} {\bibinfo
  {journal} {Nature Physics}\ }\textbf {\bibinfo {volume} {13}},\ \bibinfo
  {pages} {876–881} (\bibinfo {year} {2017})}\BibitemShut {NoStop}%
\bibitem [{\citenamefont {Tosi}\ \emph {et~al.}(2019)\citenamefont {Tosi},
  \citenamefont {Metzger}, \citenamefont {Goffman}, \citenamefont {Urbina},
  \citenamefont {Pothier}, \citenamefont {Park}, \citenamefont {Yeyati},
  \citenamefont {Nyg\aa{}rd},\ and\ \citenamefont {Krogstrup}}]{tosi-urbina}%
  \BibitemOpen
  \bibfield  {author} {\bibinfo {author} {\bibfnamefont {L.}~\bibnamefont
  {Tosi}}, \bibinfo {author} {\bibfnamefont {C.}~\bibnamefont {Metzger}},
  \bibinfo {author} {\bibfnamefont {M.~F.}\ \bibnamefont {Goffman}}, \bibinfo
  {author} {\bibfnamefont {C.}~\bibnamefont {Urbina}}, \bibinfo {author}
  {\bibfnamefont {H.}~\bibnamefont {Pothier}}, \bibinfo {author} {\bibfnamefont
  {S.}~\bibnamefont {Park}}, \bibinfo {author} {\bibfnamefont {A.~L.}\
  \bibnamefont {Yeyati}}, \bibinfo {author} {\bibfnamefont {J.}~\bibnamefont
  {Nyg\aa{}rd}},\ and\ \bibinfo {author} {\bibfnamefont {P.}~\bibnamefont
  {Krogstrup}},\ }\bibfield  {title} {\bibinfo {title} {Spin-orbit splitting of
  andreev states revealed by microwave spectroscopy},\ }\href
  {https://doi.org/10.1103/PhysRevX.9.011010} {\bibfield  {journal} {\bibinfo
  {journal} {Phys. Rev. X}\ }\textbf {\bibinfo {volume} {9}},\ \bibinfo {pages}
  {011010} (\bibinfo {year} {2019})}\BibitemShut {NoStop}%
\bibitem [{\citenamefont {Hays}\ \emph {et~al.}(2018)\citenamefont {Hays},
  \citenamefont {de~Lange}, \citenamefont {Serniak}, \citenamefont {van
  Woerkom}, \citenamefont {Bouman}, \citenamefont {Krogstrup}, \citenamefont
  {Nyg\aa{}rd}, \citenamefont {Geresdi},\ and\ \citenamefont {Devoret}}]{dev}%
  \BibitemOpen
  \bibfield  {author} {\bibinfo {author} {\bibfnamefont {M.}~\bibnamefont
  {Hays}}, \bibinfo {author} {\bibfnamefont {G.}~\bibnamefont {de~Lange}},
  \bibinfo {author} {\bibfnamefont {K.}~\bibnamefont {Serniak}}, \bibinfo
  {author} {\bibfnamefont {D.~J.}\ \bibnamefont {van Woerkom}}, \bibinfo
  {author} {\bibfnamefont {D.}~\bibnamefont {Bouman}}, \bibinfo {author}
  {\bibfnamefont {P.}~\bibnamefont {Krogstrup}}, \bibinfo {author}
  {\bibfnamefont {J.}~\bibnamefont {Nyg\aa{}rd}}, \bibinfo {author}
  {\bibfnamefont {A.}~\bibnamefont {Geresdi}},\ and\ \bibinfo {author}
  {\bibfnamefont {M.~H.}\ \bibnamefont {Devoret}},\ }\bibfield  {title}
  {\bibinfo {title} {Direct microwave measurement of andreev-bound-state
  dynamics in a semiconductor-nanowire josephson junction},\ }\href
  {https://doi.org/10.1103/PhysRevLett.121.047001} {\bibfield  {journal}
  {\bibinfo  {journal} {Phys. Rev. Lett.}\ }\textbf {\bibinfo {volume} {121}},\
  \bibinfo {pages} {047001} (\bibinfo {year} {2018})}\BibitemShut {NoStop}%
\bibitem [{\citenamefont {Peng}\ \emph {et~al.}(2015)\citenamefont {Peng},
  \citenamefont {Pientka}, \citenamefont {Vinkler-Aviv}, \citenamefont
  {Glazman},\ and\ \citenamefont {von Oppen}}]{glazman-oppen}%
  \BibitemOpen
  \bibfield  {author} {\bibinfo {author} {\bibfnamefont {Y.}~\bibnamefont
  {Peng}}, \bibinfo {author} {\bibfnamefont {F.}~\bibnamefont {Pientka}},
  \bibinfo {author} {\bibfnamefont {Y.}~\bibnamefont {Vinkler-Aviv}}, \bibinfo
  {author} {\bibfnamefont {L.~I.}\ \bibnamefont {Glazman}},\ and\ \bibinfo
  {author} {\bibfnamefont {F.}~\bibnamefont {von Oppen}},\ }\bibfield  {title}
  {\bibinfo {title} {Robust majorana conductance peaks for a superconducting
  lead},\ }\href {https://doi.org/10.1103/PhysRevLett.115.266804} {\bibfield
  {journal} {\bibinfo  {journal} {Phys. Rev. Lett.}\ }\textbf {\bibinfo
  {volume} {115}},\ \bibinfo {pages} {266804} (\bibinfo {year}
  {2015})}\BibitemShut {NoStop}%
\bibitem [{\citenamefont {Haim}\ \emph {et~al.}(2014)\citenamefont {Haim},
  \citenamefont {Keselman}, \citenamefont {Berg},\ and\ \citenamefont
  {Oreg}}]{PhysRevB.89.220504}%
  \BibitemOpen
  \bibfield  {author} {\bibinfo {author} {\bibfnamefont {A.}~\bibnamefont
  {Haim}}, \bibinfo {author} {\bibfnamefont {A.}~\bibnamefont {Keselman}},
  \bibinfo {author} {\bibfnamefont {E.}~\bibnamefont {Berg}},\ and\ \bibinfo
  {author} {\bibfnamefont {Y.}~\bibnamefont {Oreg}},\ }\bibfield  {title}
  {\bibinfo {title} {Time-reversal-invariant topological superconductivity
  induced by repulsive interactions in quantum wires},\ }\href
  {https://doi.org/10.1103/PhysRevB.89.220504} {\bibfield  {journal} {\bibinfo
  {journal} {Phys. Rev. B}\ }\textbf {\bibinfo {volume} {89}},\ \bibinfo
  {pages} {220504} (\bibinfo {year} {2014})}\BibitemShut {NoStop}%
\end{thebibliography}%

\appendix

\begin{widetext}

\section*{Computation of the BDI invariant}
\label{app}
	
Following the treatment of Ref.~\onlinecite{tudor15}, we can show that the system in Sec.~\ref{swave} decouples into independent channels with modified chemical potentials. To see this, we write the Hamiltonian in the chain index basis as:
	\begin{align}
	H_{\text{chain}}  =
	\left[ {\begin{array}{ccc}
		H_0 & -t_y\,\sigma_0 \, \tau_z\ & 0\\
		-t_y \, \sigma_0\, \tau_z\ & H_0 & -t_y\,\sigma_0\, \tau_z\\\
		0 &-t_y\, \sigma_0\,\tau_z\ & H_0
		\end{array} } \right]\,,
	\end{align}	
	where $H_0$ represents the single chain Hamiltonians. It is to be noted that each of the elements in the matrix above are themselves $4N_x\times 4N_x$ matrices. The Hamiltonian can be rewritten as:
\begin{align}
H_{\text{chain}}  = H_0 + 
\left[ {\begin{pmatrix}
0 & -t_y\,\sigma_0\,\tau_z  & 0\\
-t_y\,\sigma_0\, \tau_z & 0 & -t_y\,\sigma_0\,\tau_z\\
0 &-t_y\,\sigma_0\,\tau_z & 0
\end{pmatrix} } \right], 
\label{hchain-arnab}	
\end{align}
Rotating the Hamiltonian by the unitary matrix
$$U = \sigma_0 \,\tau_z
	\left[\begin{pmatrix}
	-1 &  1& 1 \\
	0 & \sqrt{2} & -\sqrt{2} \\
	1 &1 & 1 \\
	\end{pmatrix}\right],$$
where $U$ is constructed using the eigenvectors of the second term in Eq.~\ref{hchain-arnab}, decouples it into three sectors with modified chemical potentials of $(\mu' \pm \sqrt{2}\,t_y, \mu')$. Each of these sectors has a chiral symmetry operator of the form $\sigma_y\, \tau_y$, which can be used to define an invariant	
\begin{align}
\mathcal{Z}=\frac{1}{2\,\pi \,i}\,\int_{-\pi}^{\pi}dk\, \frac{d}{dk} \, 
\ln{ \Big [-4\, \alpha^{2}\sin^{2}k \,+\,(\mu'\,+\,2\,t_x \cos k)^2
-\left(V_x^2-\Delta_s^2 \right )\,+\,4\,i\,\Delta_s\, \alpha \sin k \Big ]}\,,
\end{align} 
for each sector, where $\mu'=\mu-2 \left (t_x+t_y  \right)$. These sectors belong to class BDI with chiral symmetry, $\mathcal{S} = \sigma_y\, \tau_y$. Fig.~\ref{1dnano}(b) exhibits the sum of the invariants for the three sectors.

\end{widetext}
	
\end{document}